\documentclass[prb,aps,twocolumn,amsmath,amssymb,floatfix,
superscriptaddress]{revtex4-1}
\usepackage[dvips]{graphics}
\usepackage{color}
\definecolor{dred}{rgb}{0,0,0.6}
\usepackage{graphicx}  
\usepackage{dcolumn}   
\usepackage{bm}        
\usepackage{amssymb}   
\usepackage{amsmath}
\usepackage{braket}
\usepackage[mathscr]{euscript}
\usepackage{color}
\usepackage{hyperref}
\usepackage{multirow}
\usepackage{ragged2e}
\begin{document}

\title{Persistent current in a non-Hermitian Hatano-Nelson ring: Disorder-induced amplification}

\author{Sudin Ganguly}

\email{sudinganguly@gmail.com}

\affiliation{Department of Physics, School of Applied Sciences, University of Science and Technology Meghalaya, Ri-Bhoi-793 101, India}

\author{Santanu K. Maiti}

\email{santanu.maiti@isical.ac.in}

\affiliation{Physics and Applied Mathematics Unit, Indian Statistical
  Institute, 203 Barrackpore Trunk Road, Kolkata-700 108, India}


\begin{abstract}
Non-reciprocal hopping induces a synthetic magnetic flux which leads to the non-Hermitian Aharonov-Bohm effect. Since non-Hermitian Hamiltonians possess both real and imaginary eigenvalues, this effect allows the observation of real and imaginary persistent currents in a ring threaded by the synthetic flux. Motivated by this, we investigate the behavior of persistent currents in a disordered Hatano-Nelson ring with anti-Hermitian intradimer hopping. The disorder is diagonal and we explore three distinct models, namely the Aubry-Andr\'{e}-Harper model, the Fibonacci model, both representing correlated disorder, and an uncorrelated (random) model. We conduct a detailed analysis of the energy spectrum and examine the real and imaginary parts of the persistent current under various conditions such as different ring sizes and filling factors. Interestingly, we find that real and imaginary persistent currents exhibit amplification in the presence of correlated disorder. This amplification is also observed in certain individual random configurations but vanishes after configuration averaging. Additionally, we observe both diamagnetic and paramagnetic responses in the current behavior and investigate aspects of persistent currents in the absence of disorder that have not been previously explored. Interestingly, we find that the intradimer bonds host only imaginary currents, while the interdimer bonds carry only real currents. The bulk-boundary correspondence is investigated by analyzing the existence of localized edge states under the open boundary condition.
\end{abstract}

\maketitle
\section{\label{sec1}Introduction}
Persistent current arises in a metal ring when electron mean free path exceeds ring circumference, enclosing magnetic flux. It was first predicted by Kulik~\cite{pc1} and later by B\"{u}ttiker, Imry, and Landauer in the context of a 1D disordered ring~\cite{pc2}. A few years later, the existence of persistent current was further affirmed by notable experimental investigations conducted on $10^7$ isolated mesoscopic copper rings~\cite{pcex1} and an isolated gold ring~\cite{pcex2}. So far, numerous studies have been conducted to explore persistent currents in various types of mesoscopic rings under different scenarios, both theoretically~\cite{pc3,pc4,pc5,pc6,pc7,pc8,pc9,pc10,pc11} and experimentally~\cite{pcex3,pcex4,pcex5,pcex6}.

In the aforementioned studies, it is noteworthy that all theoretical investigations are grounded in Hermitian models. Another crucial aspect of persistent currents is their reliance on a real electromagnetic potential within closed-loop structures. However, two seminal works by Hatano and Nelson showed that asymmetric coupling can induce a synthetic gauge field, which also makes the system non-Hermitian~\cite{hatano1,hatano2}. Consequently, in recent years, several attempts have been made to study the effect of synthetic gauge fields utilizing non-reciprocal hopping~\cite{nrh1,nrh2,nrh3,nrh4,nrh5,nrh6,nrh7,nrh8}. For instance, studies have examined 1D tight-binding lattices to explore transport properties~\cite{nrh1}, transport properties of light in 1D photonic lattices~\cite{nrh2}, Bloch oscillations in the non-Hermitian Hatano-Nelson (HN) model~\cite{nrh3}, 1D chains of noninteracting bosonic cavities~\cite{nrh5}, etc. In Ref.~\cite{nrh8}, it has been demonstrated that a magnetic flux can be achieved in a non-Hermitian Hatano-Nelson ring by incorporating anti-Hermitian intradimer hopping. In presence of this flux, both real and imaginary persistent currents can be generated.

Inspired by recent work~\cite{nrh8}, we explore the behavior of persistent current in an HN ring with anti-Hermitian intradimer hopping under different disordered scenario. Specifically, three different types of disorder is considered, namely, the well-known Aubry-Andr\'{e}-Harper (AAH) model~\cite{aah1,aah2}, Fibonacci~\cite{fb1,fb2}, and uncorrelated (random) disorder~\cite{random}. Additionally, we investigate several aspects in the absence of disorder that have not been previously discussed, specifically the characteristic features of the energy spectrum and persistent currents with varying ring sizes, and the influence of the filling factor. We further investigate the nature of the persistent current across different bonds of the ring. To gain insights into the topological and trivial phases, we investigate the behavior of localized edge modes under open boundary condition. This analysis is conducted both in the absence and presence of the three aforementioned disorder types. We also present a proposal for experimental feasibility, outlining how to implement anti-Hermitian hopping, various types of disorder, and other key elements.

The key results of the present work are: 
(i) while the real (imaginary) persistent currents display a notable increase in magnitude within the trivial (topological) regime, the smaller component of currents does not vanish entirely,
(ii) correlated disorder induces anomalous signatures in the energy spectra, 
(iii) the presence of correlated disorder enhances the magnitude of real and imaginary persistent currents compared to the disorder-free scenario,
(iv) tuning the synthetic magnetic flux allows for the realization of both diamagnetic and paramagnetic responses multiple times within a single flux period, surpassing the behavior observed in the disorder-free case, and
(v) the persistent current is always imaginary in the intra-dimer bonds, with the real current being identically zero, while it is always real in the inter-dimer bonds, with the imaginary current being identically zero.

The rest of the paper is organized as follows. In the next section (Section~\ref{sec2}), we present the schematic diagram, model Hamiltonian, and the necessary theory to compute the persistent current. In Section~\ref{sec3}, we discuss the results, including the energy spectrum, ground state energy, and real and imaginary persistent currents under different scenarios both in the absence and presence of disorder. We also include a detailed derivation for the bond-resolved persistent currents along with their features. The key results, both in the absence and presence of disorder, are summarized in tabular form along with the experimental proposal.  Finally, we summarize our findings in Section~\ref{conclusion}.

\section{\label{sec2}Physical system, Hamiltonian and Theoretical Framework}
\begin{figure}[h] 
\includegraphics[width=0.4\textwidth]{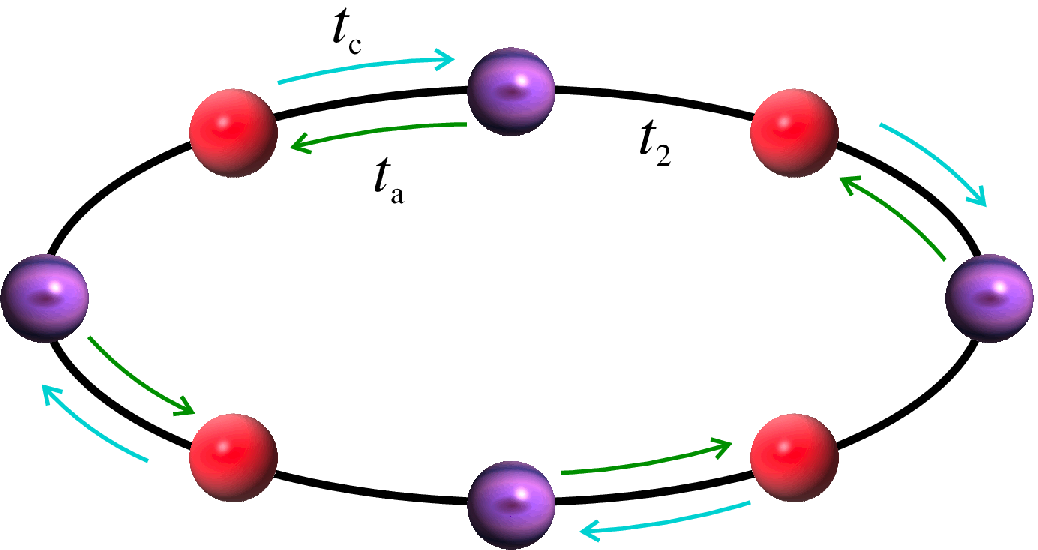}
\caption{(Color online.) A schematic representation of a Hatano-Nelson ring consisting of eight sites is shown with two sublattices labeled, sublattice $A$ in red and sublattice $B$ in violet. The intradimer hopping strength in the clockwise direction is $t_c$, shown in cyan, and in the counterclockwise direction is $t_a$. The interdimer hopping strength is denoted by $t_2$.}
\label{setup}
\end{figure}
The schematic representation of an HN ring is shown in Fig.~\ref{setup}. The lattice structure is composed of two sites per unit cell. The asymmetry is introduced in the intradimer bond. The corresponding nearest-neighbor tight-binding Hamiltonian is described as
\begin{eqnarray}
H &=& \sum\limits_{n=1}^{N} \left(\epsilon_{n,A} c_{n,A}^{\dagger} c_{n,A} + \epsilon_{n,B} c_{n,B}^{\dagger} c_{n,B}\right)\nonumber\\&+&
\sum\limits_{n=1}^N\left(t_{c} c_{n,A}^{\dagger} c_{n,B} + t_{a} c_{n,B}^{\dagger} c_{n,A}\right)  \nonumber\\&+&\sum\limits_{n=1}^{N-1}\left( t_{2} c_{n,B}^{\dagger} c_{n+1,A} + h.c\right) 
\nonumber\\&+&
\left( t_2 c^\dagger_{N,B} c_{1,A} + h.c.\right),
\label{ham}
\end{eqnarray}
where $N$ is the number of unit cells in the HN ring. $A$ and $B$ are the two sublattices. $c_{n,\alpha}^\dagger\left(c_{n,\alpha}\right)$ is the creation (annihilation) operator in the $n$-th unit cell in the sublattice-$\alpha~(=A,B)$.  The first and second terms in Eq.~\ref{ham} represent the on-site terms, where $\epsilon_{n,A}$ and $\epsilon_{n,B}$ are the on-site energies at the $n$-th unit cell in the $A$ and $B$ sublattices, respectively. The third term is the non-reciprocal intracell hopping term. $t_c$ is the hopping strength from sublattice-$A$ to sublattice-$B$ (clockwise sense) and $t_a$ is the hopping strength from sublattice-$B$ to sublattice-$A$ (anticlockwise sense) in the same unit cell. The non-hermiticity is introduced in the system through the intracell hopping and $t_c=-t_a^*$, that is the intracell hopping is antihermitian. The fourth term is the intercell hopping term, where $t_2$ is the intercell hopping strength. The fifth term connects the first and $N$-th unit cells through the intercell hopping strength $t_2$. 

The disorder in the ring is introduced through the on-site energies. In this work, we consider three different types of disorder: (i) the Aubry-Andr\'{e}-Harper (AAH) model~\cite{aah1,aah2}, (ii) the Fibonacci model~\cite{fb1,fb2}, and (iii) a random distribution~\cite{random}. The first two models are well-known correlated disorder, while the third model introduces uncorrelated disorder, as the name suggests. 

The AAH model has the functional form  for the $A$ sublattice as $\epsilon_{n,A}=W\cos{[2\pi b (2n-1)]}$ and for $B$-sublattice $\epsilon_{n,B}=W\cos{[2\pi b (2n)]}$. Here $W$ is the strength of disorder. $b$ is an irrational number and considered as~\cite{aah1} $b=(1+\sqrt{5})/2$. 

The Fibonacci sequence can be obtained iteratively with the modified relation $S_{n} = S_{n-1}S_{n-2}$ and the initial conditions $S_0=Y$ and $S_1=X$. Using these definitions, the first few sequences are $Y, X, XY, XYX, XYXXY, XYXXYXYX$. $X$ and $Y$ are treated as two basic units of the ring. To specify the on-site potentials based on this sequence, we assign values to the potentials according to whether the site unit is $X$ or $Y$. Specifically, when the site unit is $X$, $\epsilon_{n,A/B}=W$ and when it is $Y$, $\epsilon_{n,A/B}=-W$.

For the uncorrelated disorder, all the on-site potentials will be picked up randomly from a box distribution between $-W$ to $W$.  

The intracell hopping can be considered as $t_c=t+i\gamma$ and $t_a=-t+i\gamma$, where both $t$ and $\gamma$ are real numbers. They can be written in polar form as $t_c=\lvert t_1 \rvert e^{i\phi}$ and $t_a=-\lvert t_1 \rvert e^{-i\phi}$, where the hopping amplitude $\lvert t_1\rvert=\sqrt{t^2 + \gamma^2}$ and $e^{i\phi}$ is the Peierls phase factor. Over a complete cycle the total phase acquired is $e^{iN\phi}$ and $\Phi=N\phi$, where $\Phi$ can be identified as a real magnetic field generated by the non-Hermitian system~\cite{nrh8}. In our system, the interdimer hopping strength $t_2$ is always real.

At absolute zero temperature, electrons in a system will occupy the energy levels starting from the lowest available level, filling them sequentially in accordance with the Pauli exclusion principle for the spin-less case. Since the Hamiltonian of the system is non-Hermitian, its eigenvalues extend over both real and imaginary spaces. As a result, the real and imaginary Fermi surfaces are defined separately, each comprising only the real or imaginary eigenvalues, respectively. Consequently, the ground state energies are computed independently in the real and imaginary energy spaces. The persistent current can be determined using two approaches: (i) by computing the ground state energy and taking its derivative with respect to the flux $\Phi$ or (ii) by calculating the current contribution from each individual state and summing over all occupied energy levels based on the filling factor. The detailed steps for these calculations are given below. 

(i) {\it From the ground state energies}: 
If the number of electrons in the system is $N_e$, and this corresponds to the index $m$ of the highest occupied energy level, then the ground state energies of the system can be determined by summing the energies of all occupied levels up to $m$ as
\begin{eqnarray}
\text{Re}\left[E_G\right] &=& \sum_{n=1}^m \text{Re}\left[E_n\right], \\
\text{Im}\left[E_G\right] &=& \sum_{n=1}^m \text{Im}\left[E_n\right],
\end{eqnarray}
where $m$ denotes the highest occupied energy level. Correspondingly, the real and imaginary persistent currents are defined as~\cite{nrh8}
\begin{eqnarray}
\text{Re}[I] &=&  -c\frac{\partial}{\partial \Phi}\text{Re}\left[E_G\right],\label{step11}\\
\text{Im}[I] &=&  -c\frac{\partial}{\partial \Phi}\text{Im}\left[E_G\right],\label{step12}
\end{eqnarray}
where $c$ is the speed of light in free space. 

(ii) {\it From the individual state}:
The real and imaginary persistent currents for the $n$-th state are defined as~\cite{nrh8}
\begin{eqnarray}
\text{Re}\left[I_n\right] &=& -c\frac{\partial}{\partial \Phi}\text{Re}\left[E_n\right],\\
\text{Im}\left[I_n\right] &=& -c\frac{\partial}{\partial \Phi}\text{Im}\left[E_n\right],
\end{eqnarray} 
where $c$ is the speed of light in free space.  The net real and imaginary persistent currents by the occupied electronic states will be then
\begin{eqnarray}
\text{Re}[I] &=& \sum_{n=1}^m \text{Re}\left[I_n\right] \label{step21},\\
\text{Im}[I] &=& \sum_{n=1}^m \text{Im}\left[I_n\right].\label{step22}
\end{eqnarray}

Both approaches ultimately lead to the same result. Specifically, Eqs.~\ref{step11} and \ref{step21} yield identical values for the real current, while Eqs.~\ref{step12} and \ref{step22} produce the same result for the imaginary current.

\section{\label{sec3}Results and Discussion}
Throughout the discussion, all energies are measured in units of eV. The intercell hopping is fixed at $t_2=1\,$eV. We first consider the case without disorder and discuss several important aspects that have not been explored earlier. 
\begin{figure*}[ht] 
\includegraphics[width=0.3\textwidth]{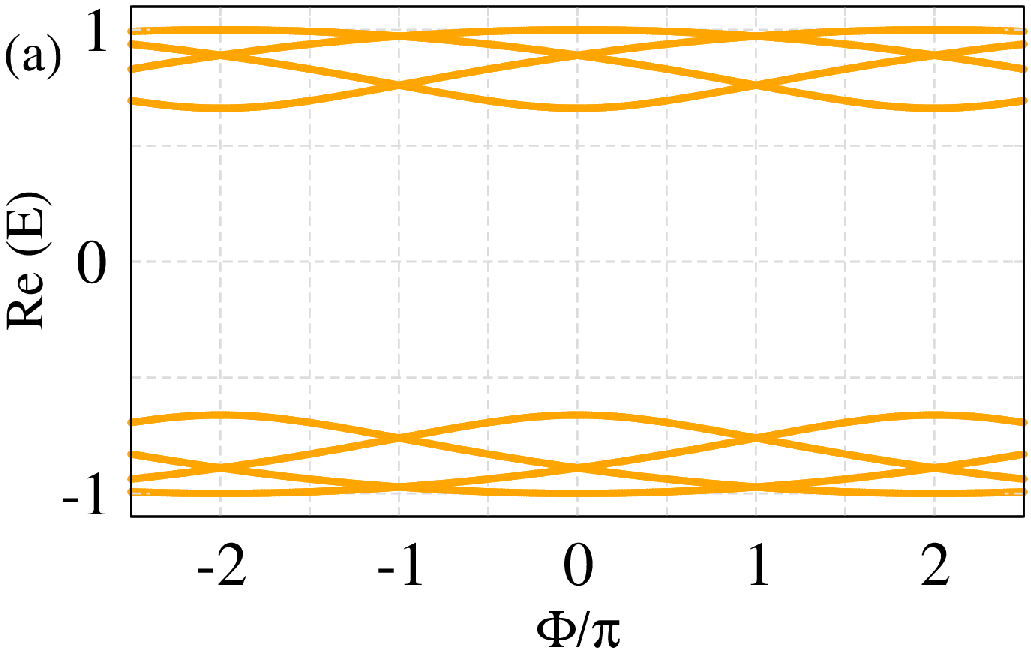}\hfill
\includegraphics[width=0.3\textwidth]{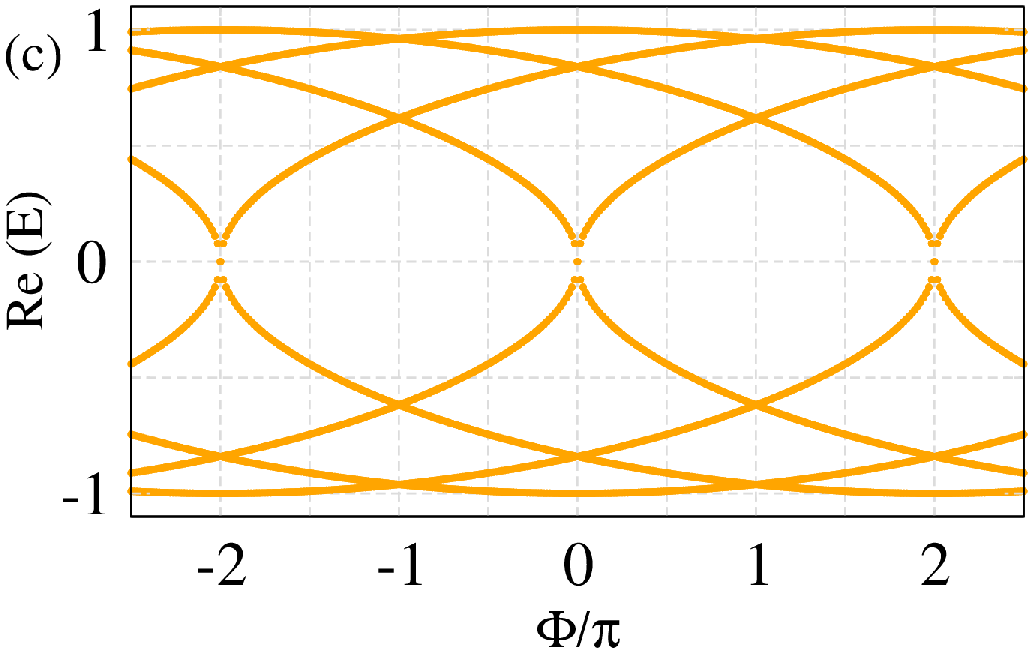}\hfill
\includegraphics[width=0.3\textwidth]{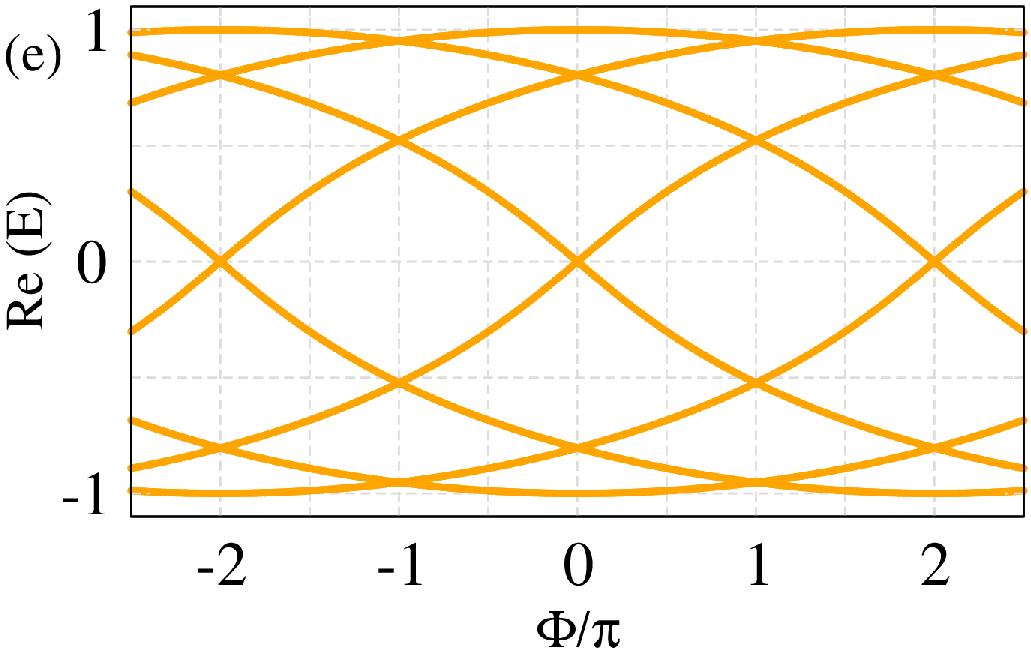}\vskip 0.1 in
\includegraphics[width=0.3\textwidth]{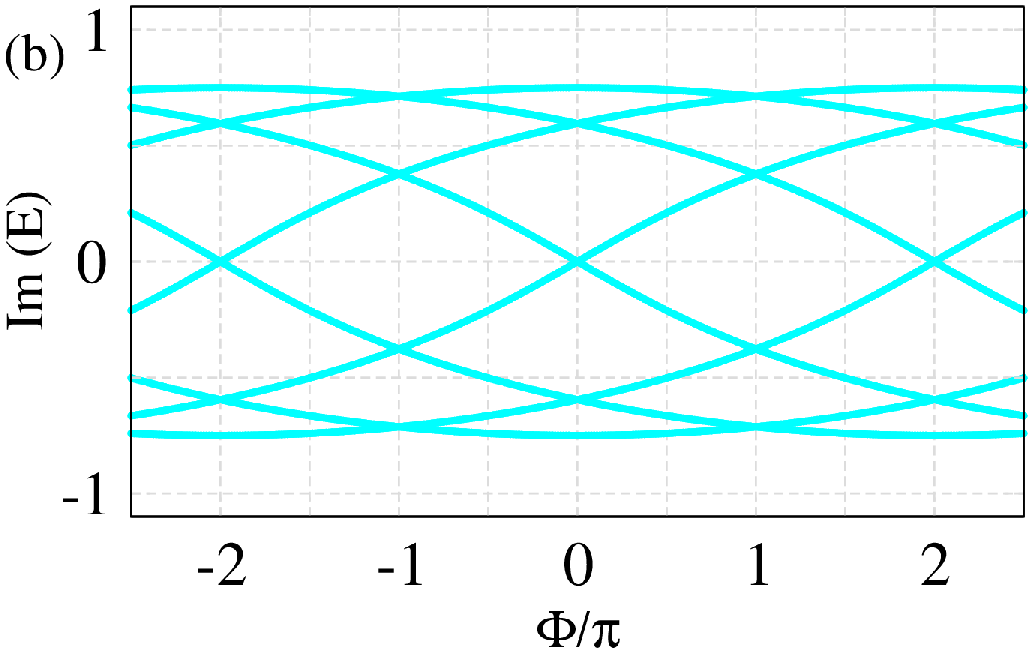}\hfill
\includegraphics[width=0.3\textwidth]{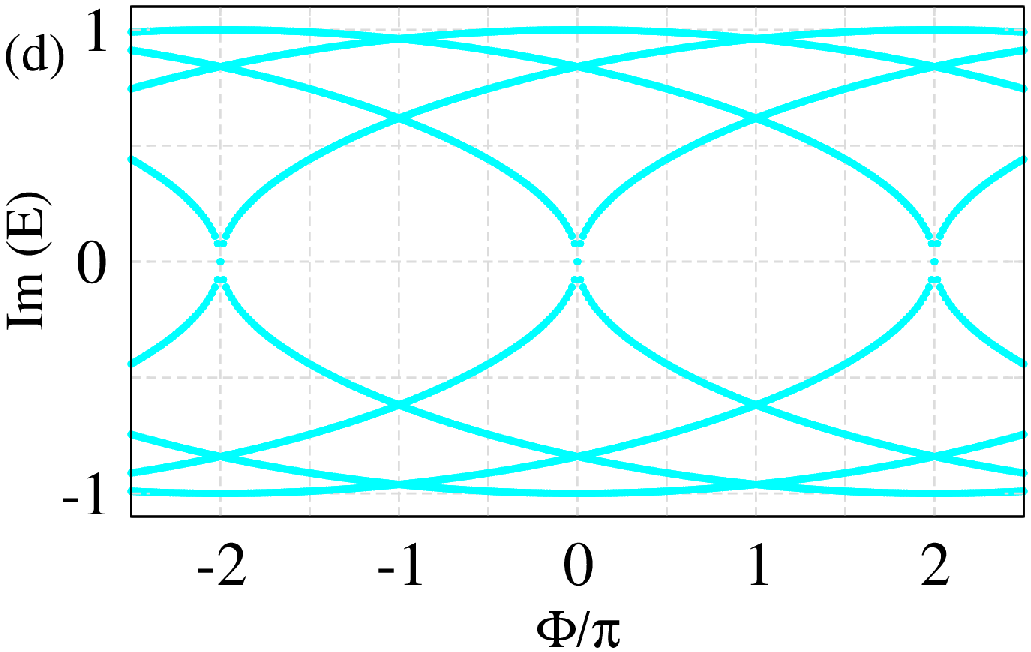}\hfill
\includegraphics[width=0.3\textwidth]{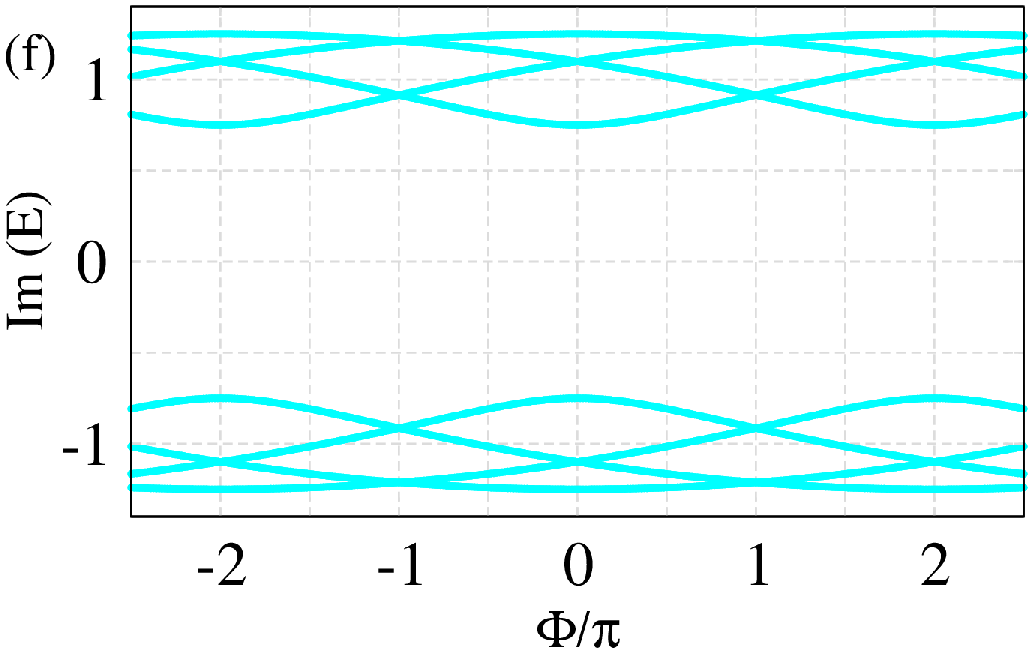}
\caption{(Color online.) Real and imaginary energy eigenspectra as a function of $\Phi$, where the first, second and, third columns are for $t_1=0.75$, $1$, and $1.25$, respectively. Here, the number of unit cells $N=8$ and the intercell hopping strength $t_2=1$. }
\label{ephi}
\end{figure*}
\subsection{Absence of disorder}
First, we study the behavior of the complex energy spectrum as a function of $\Phi$ as shown in Fig.~\ref{ephi}. We consider the number of unit cells $N=8$. Three different cases are considered here, namely, the topological phase $t_1 < t_2$, trivial phase $t_1 > t_2$, and the transition point from topological to trivial phase $t_1 = t_2$. The explicit nature of these phases in the context of topological invariant is discussed latter in the appropriate sub-section. 

In the topological phase $(t_1<t_2)$, the real energy spectrum exhibits a gap which was identified as a real line gap in an earlier study~\cite{nrh8} as shown in Fig.~\ref{ephi}(a). A real (imaginary) line gap exists if the ${\mathbf k}$-space non-Hermitian Hamiltonian is invertible for all ${\mathbf k}$ values and all the real (imaginary) eigenvalues are finite~\cite{prx}. The imaginary eigenvalues continuously evolve with $\Phi$ and the spectrum is gapless as depicted in Fig.~\ref{ephi}(b). Both the real and imaginary energy spectra oscillate with $\Phi$ with a period $2\pi$.

At the transition point $(t_1=t_2)$, the non-Hermitian Hamiltonian yields the same set of real and imaginary eigenvalues. This is evident from the expression of the $n$-th eigenenergy $E_n = \pm\sqrt{t_2^2 - \lvert t_1 \rvert^2 + 2 \lvert t_1\rvert t_2 i \sin{\left[\frac{2\pi}{N}\left(n+\Phi/\Phi_0\right)\right]}}$, where $\Phi_0$ is the magnetic flux-quantum. Consequently, the real and imaginary eigenvalues are identical as a function of $\Phi$ as shown in Figs.~\ref{ephi}(c) and (d), respectively. Both the spectra exhibit gapless band structure.

The trivial phase $(t_1>t_2)$ presents a completely opposite scenario compared to the topological phase. The real energy values continuously evolve with $\Phi$ and the corresponding spectrum is gapless as depicted in Fig.~\ref{ephi}(e). On the other hand, the imaginary spectrum is associated with a gap, which, according to the criteria, is identified as a line gap. A detailed analysis of all these spectrum can be found in Ref.~\cite{nrh8}.

Once we have the energy spectrum as a function of $\Phi$, we are ready to analyze the behavior of the persistent current. Since persistent current is essentially the slope of the ground state energy, let us first examine how the ground state energy behaves with the magnetic flux, considering both its real and imaginary parts. To compute the ground state energy $E_G$, we consider a relatively bigger ring size with number of unit cells $N=20$ and fix the number of electrons $N_e=20$, which is the half-filled case. 

\begin{figure*}[ht]
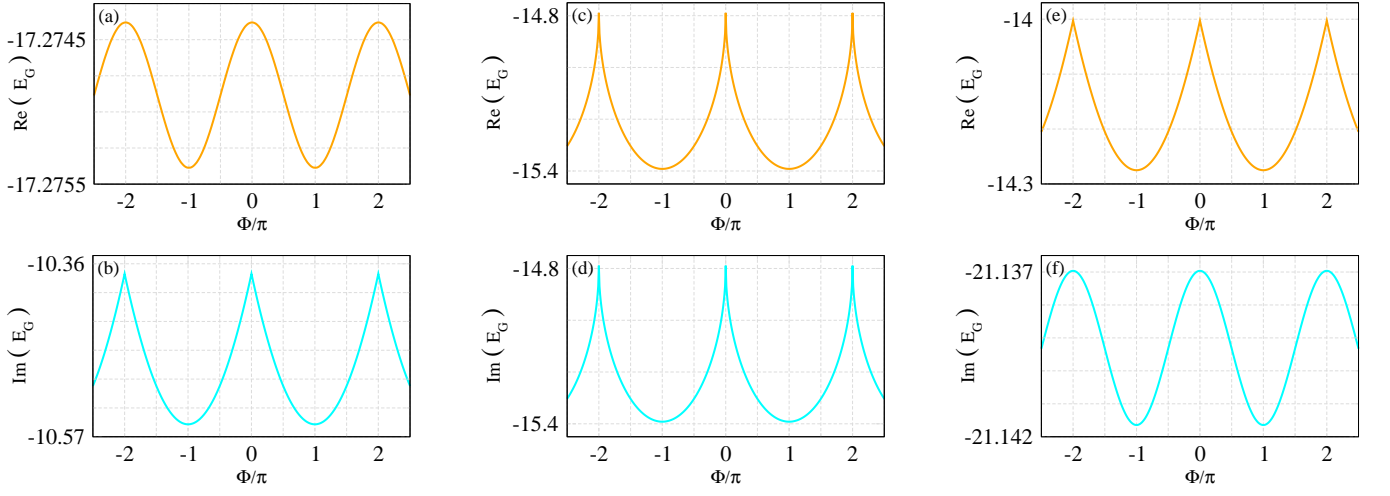
 
\includegraphics[width=0.3\textwidth]{fig3a.eps}\hfill
\includegraphics[width=0.3\textwidth]{fig3c.eps}\hfill
\includegraphics[width=0.3\textwidth]{fig3e.eps}\vskip 0.1 in
\includegraphics[width=0.3\textwidth]{fig3b.eps}\hfill
\includegraphics[width=0.3\textwidth]{fig3d.eps}\hfill
\includegraphics[width=0.3\textwidth]{fig3f.eps}
\caption{(Color online.) Ground state energy $E_G$ variation with $\Phi$, where the first, second, and third columns are for $t_1=0.75$, $1$, and $1.25$, respectively. Here, the number of unit cells $N=20$ and the intercell hopping strength $t_2=1$.  Number of electrons is $N_e = 20$.}
\label{gs}
\end{figure*}

In the topological phase $(t_1<t_2)$, the real and imaginary parts of the ground state energy are depicted in Figs.~\ref{gs}(a) and (b), respectively. The real part of $E_G$ shows sinusoidal behavior with $\Phi$. The difference between the maximum and minimum magnitudes of Re$\left[ E_G\right]$ is about $10^{-3}$ and this amount is spread over a full cycle of oscillation, that is $2\pi$. Therefore, it is clear that the slope of the ground state energy would be very less, which consequently will reflect on the behavior of the current as we shall see. The imaginary part of the ground state energy, however, exhibits a periodic, parabola-like pattern with $\Phi$, which is typically encountered in simple 1D metallic rings for the half-filled case. The imaginary part of $E_G$ is discontinuous at $\Phi=0$ and $\pm 2\pi$. The same periodicity is also observed in both the real and imaginary parts of the ground state energy, similar to the energy spectrum (Fig.~\ref{ephi}).
 
\begin{figure*}[ht]
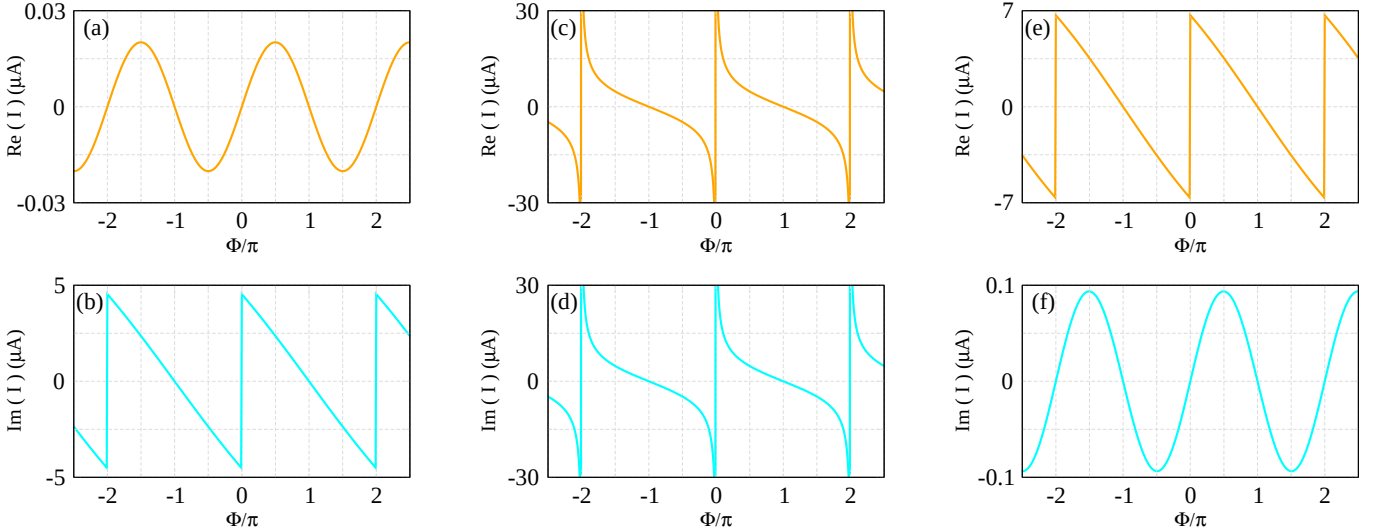
 
\includegraphics[width=0.3\textwidth]{fig4a.eps}\hfill
\includegraphics[width=0.3\textwidth]{fig4c.eps}\hfill
\includegraphics[width=0.3\textwidth]{fig4e.eps}\vskip 0.1 in
\includegraphics[width=0.3\textwidth]{fig4b.eps}\hfill
\includegraphics[width=0.3\textwidth]{fig4d.eps}\hfill
\includegraphics[width=0.3\textwidth]{fig4f.eps}
\caption{(Color online.) Persistent current $I$ versus magnetic flux $\Phi$. The first, second, and third columns are for $t_1=0.75$, $1$, and $1.25$, respectively. All the other parameters are same as described in Fig.~\ref{gs}}
\label{ip}
\end{figure*}
 
At the transition point $(t_1 = t_2)$, both the real and imaginary parts of $E_G$ are identical as a function of $\Phi$, as shown in Figs.~\ref{gs}(c) and (d), respectively. Here, the energy is also periodic with $\Phi$, displaying a parabola-like pattern. This identical behavior is evident in view of the energy spectrum at the transition point (Figs.~\ref{ephi}(c) and (d)). The $E_G$-$\Phi$ curves are found to be discontinuous at $\Phi=0$ and $\pm 2\pi$.

In the trivial phase $(t_1 > t_2)$, the behavior of the ground state energy is completely opposite to that in the topological phase. Here, the real part of $E_G$ exhibits a parabola-like pattern (Fig.~\ref{gs}(e)), while the imaginary part of $E_G$ shows sinusoidal behavior (Fig.~\ref{gs}(f)). Similar to the real part in the topological phase, for the imaginary part in the trivial phase, the difference between the maximum and minimum magnitudes of $\text{Re}[E_G]$ is about $5 \times 10^{-3}$, which is spread over a period of $2\pi$. Consequently, the slope of the energy curve is less than that for the real part. The real part of $E_G$ is also discontinuous at $\Phi=0$ and $\pm 2\pi$.

The behavior of persistent current as a function of $\Phi$ is illustrated in Fig.~\ref{ip} which we compute by taking the slope of the graphs in Fig.~\ref{gs}. All other system parameters are same as in Fig.~\ref{gs}. 

The topological phase reveals an interesting feature regarding the real part of the persistent current, which is finite. This particular point has not been explored in previous works. In Ref.~\cite{nrh8}, it was stated that the persistent current is identically zero in the topological phase. However, as is clearly seen from Fig.~\ref{ip}(a), this is not the case. The magnitude is small ($\sim 0.03\,\mu$A), as predicted from the ground state behavior, but it is finite for the considered ring size in the half-filled case. On the other hand, the imaginary current is quite large ($\sim 5\,\mu$A) compared to its real counterpart and exhibits a saw-tooth pattern similar to that in Hermitian 1D rings~\cite{pc3}. Like the ground state energy, Im$\left[I\right]$ is also discontinuous at $\Phi=0$ and $\pm 2\pi$. These are the points where Im$\left[E_G\right]$ changes the sign of the slope discontinuously. The periodicity in both cases is $2\pi$.

At the transition point, the behavior of the real and imaginary parts of the persistent current is identical as a function of $\Phi$, as shown in Figs.~\ref{ip}(c) and (d), respectively. This is expected since the energy spectra and the ground state energies are identical for both the real and imaginary parts at the transition point. The persistent current can be obtained from the expression of the $n$-th energy level and is given by~\cite{nrh8}
$$I_n = \frac{-i I_0 t_1 \cos{\phi_n}}{\sqrt{t_2^2 - t_1^2 + 2 i t_1 t_2 \sin{\phi_n}}}.$$
Where $I_0=2\pi c t_2/N\Phi_0$ and $\phi_n = \frac{2\pi}{N}\left(n+\frac{\Phi}{\Phi_0}\right)$. At $\Phi = 0$ and $\pm 2\pi$, the denominator becomes zero, as certain currents carried by their respective energy levels are undefined. For instance, the denominator becomes zero for $\phi_n=0$, which happens for the level $n=0$, and consequently, corresponding current becomes undefined. This is reflected in the $I$-$\Phi$ graphs, where both parts diverge at these points. Such divergences are quite sensitive to the number of electrons considered in the system. 

The $I$-$\Phi$ characteristics in the trivial phase are completely opposite to those in the topological phase, as shown in Figs.~\ref{ip}(e) and (f). The real (imaginary) current exhibits a similar behavior to the imaginary (real) current in the topological phase. Another important unexplored point to note here is that the imaginary current is not identically zero but finite in the trivial phase (Fig.~\ref{ip}(f)).  

In all the $I$-$\Phi$ plots, both the real and imaginary currents are zero at $\Phi=0,\pm\pi,\pm 2\pi$. This is due to the fact that at $\Phi=0,\pm 2\pi$, the ground state energy is maximum, while at $\Phi=\pm \pi$, it is minimum. Since the current is proportional to the slope of the ground state energy, we get zero current at those $\Phi$-values. These are also the points where all the currents change their sign from positive to negative or vice versa.

Overall, we find two important points from the above discussion. The real persistent current in the topological phase and the imaginary current in the trivial phase are not identically zero but finite for the considered system size and number of electrons. Now, let us see to what extent the persistent current remains non-zero as we increase the system size, knowing that with an increasing ring size, the current will eventually go to zero. 

\begin{figure}[h] 
\includegraphics[width=0.48\textwidth]{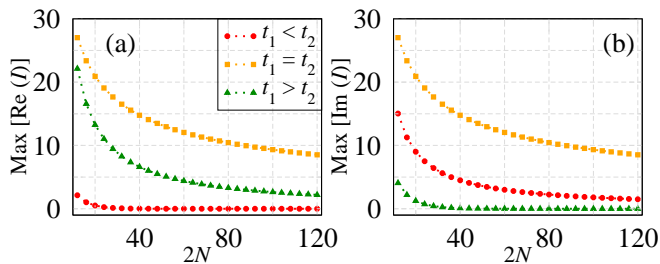}
\caption{(Color online.) Maximum persistent current versus number of sites $2N$ in the ring. (a) Real and (b) imaginary parts of the maximum current. Red, orange, and green curves represent the results corresponding to the topological phase, transition point, and trivial phase, respectively. The hopping parameters are the same as in the previous plots for the different phases. The current is computed for the half-filled case.}
\label{size}
\end{figure}

The behavior of persistent current with ring size is depicted in Fig.~\ref{size}. In Fig.~\ref{size}(a), we plot the maximum real current and in Fig.~\ref{size}(b) the maximum imaginary current as a function of $2N$, the number sites in the system. The maximum current is obtained over a complete period, that is by varying the magnetic flux $\Phi$ from 0 to $2\pi$. Here also we consider the half-filled case. The number of sites is chosen to be a multiple of 4 in order to have an even number of electrons for the half-filled case starting from $2N=8$ to 120. For the transition point case, since both the real and imaginary currents tend to diverge at $\Phi=0$ and $\pm 2\pi$, these points are excluded from the $\Phi$-range, within which the maximum currents are picked up. Both the real and imaginary parts of the persistent current exhibit the usual decrease in magnitude as we increase the ring size.
However, it is important to note that the real part in the topological phase becomes zero beyond $2N \approx 20$ (Fig.~\ref{size}(a)), while the imaginary part in the trivial phase becomes zero beyond $2N \approx 24$ (Fig.~\ref{size}(b)). On the other hand, the real part in the trivial phase and the imaginary part in the topological phase decay slowly with the ring size. Even for a large system size, they attain a finite current. For instance the maximum current is about a few $\,\mu$A for $2N=120$.

\begin{figure}[h] 
\includegraphics[width=0.48\textwidth]{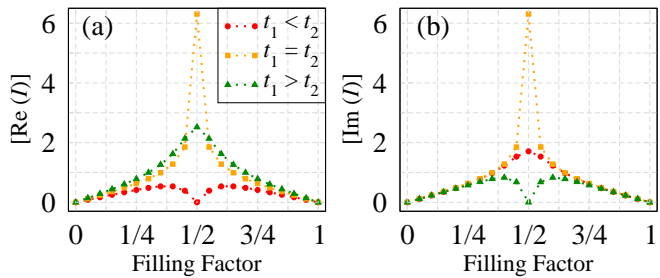}
\caption{(Color online.) Real and imaginary persistent current as a function of filling factor. (a) Real and (b) imaginary parts of the maximum current. The color schemes and the hopping parameters for different phases are the same as in Fig.~\ref{size}. The system size is considered as $N=40$. The flux is fixed at $\Phi=\pi/4$.}
\label{fill}
\end{figure}
\begin{figure*}[ht] 
\includegraphics[width=0.3\textwidth]{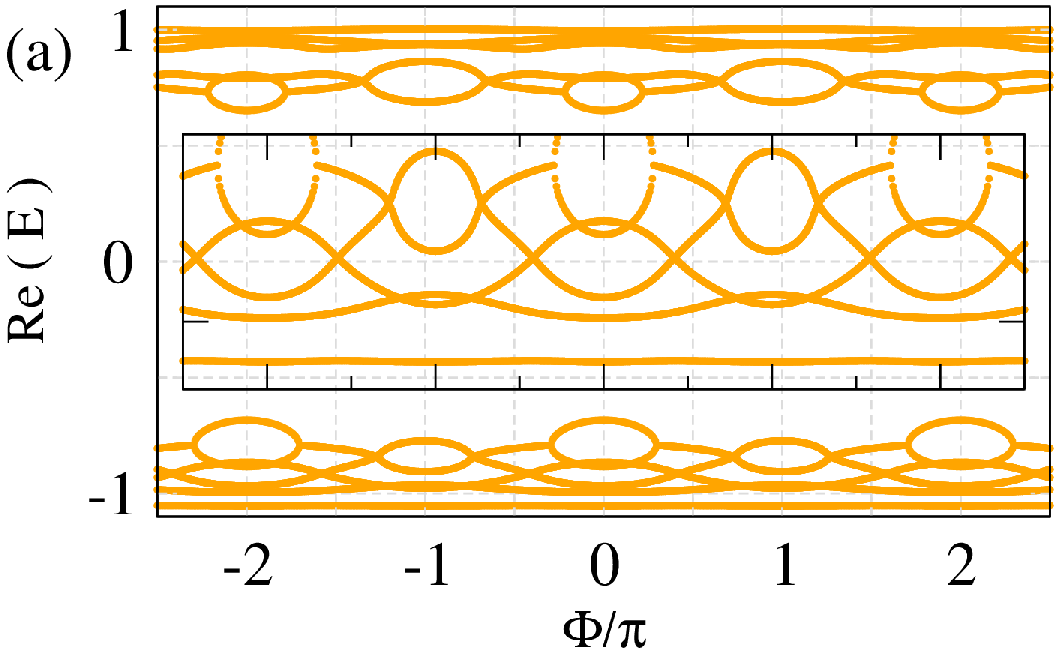}\hfill
\includegraphics[width=0.3\textwidth]{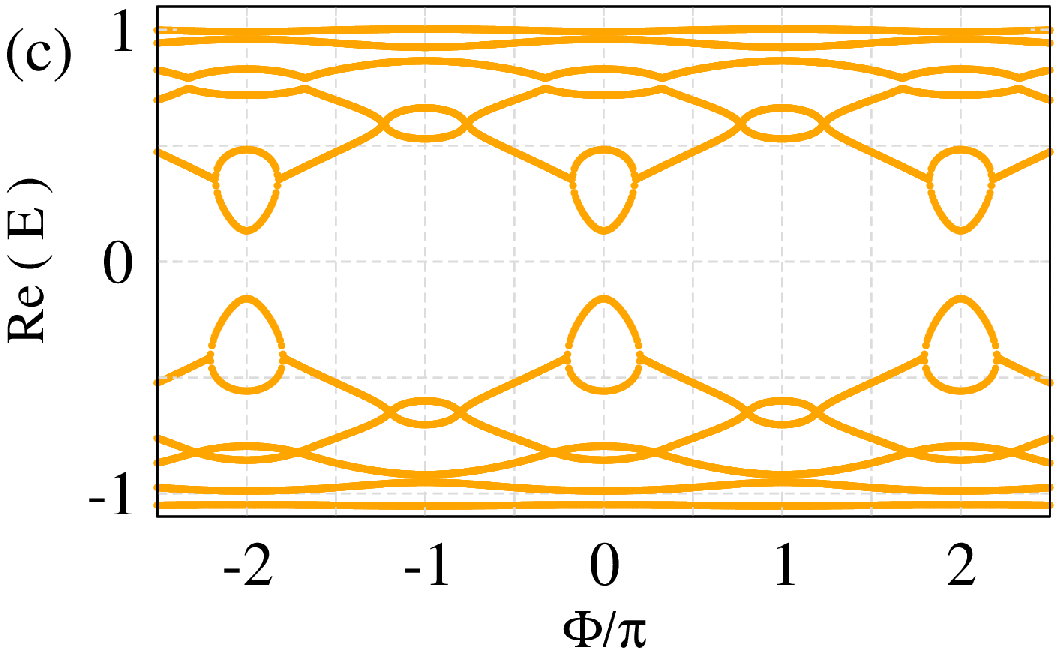}\hfill
\includegraphics[width=0.3\textwidth]{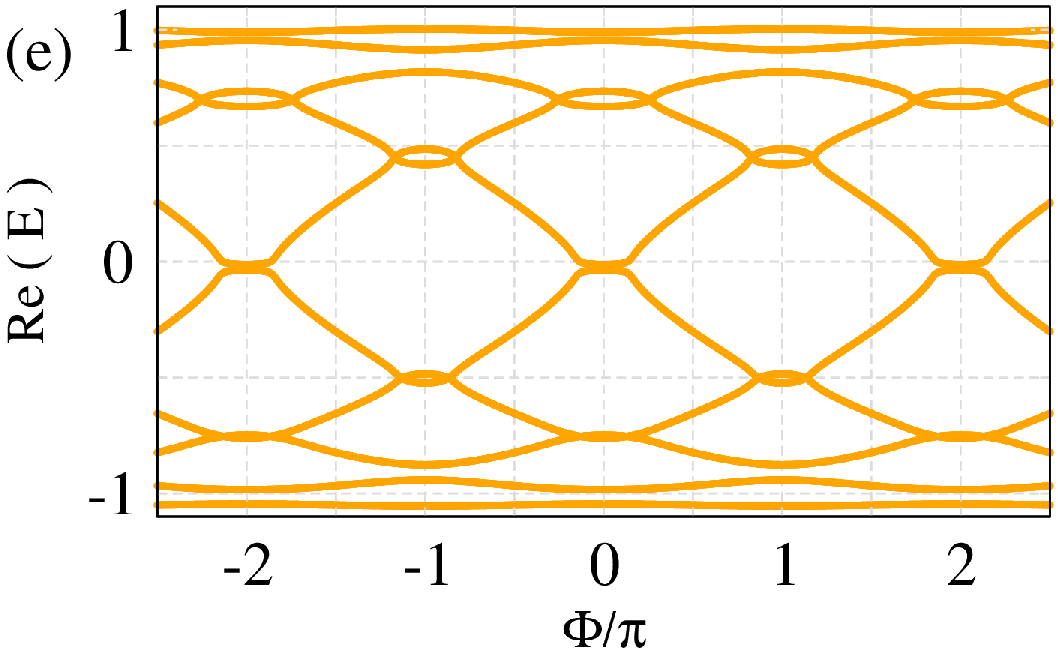}\vskip 0.1 in
\includegraphics[width=0.3\textwidth]{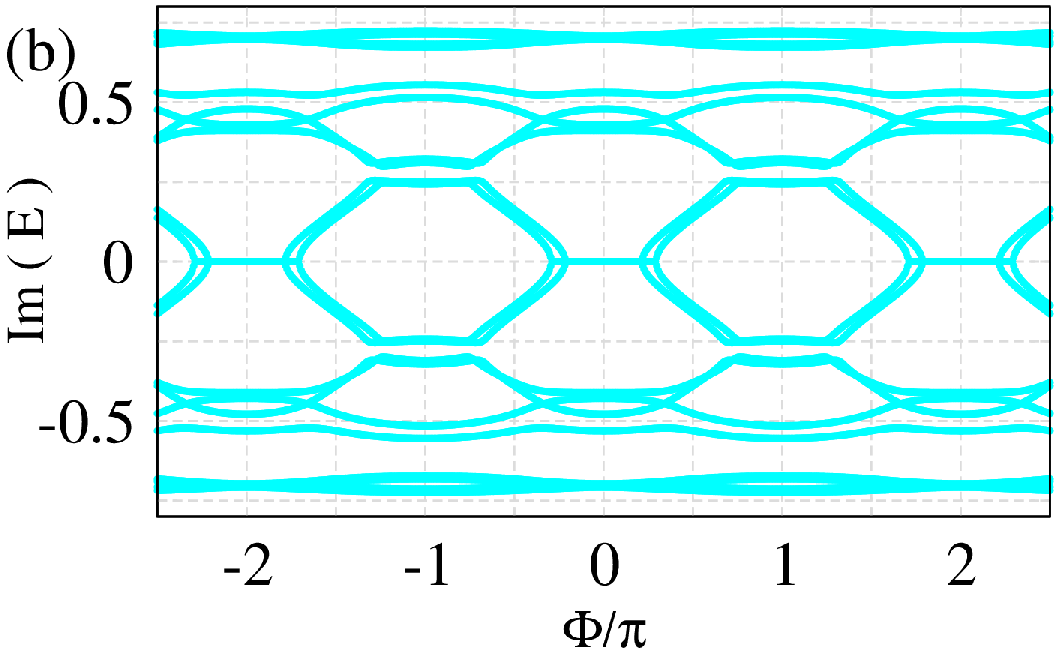}\hfill
\includegraphics[width=0.3\textwidth]{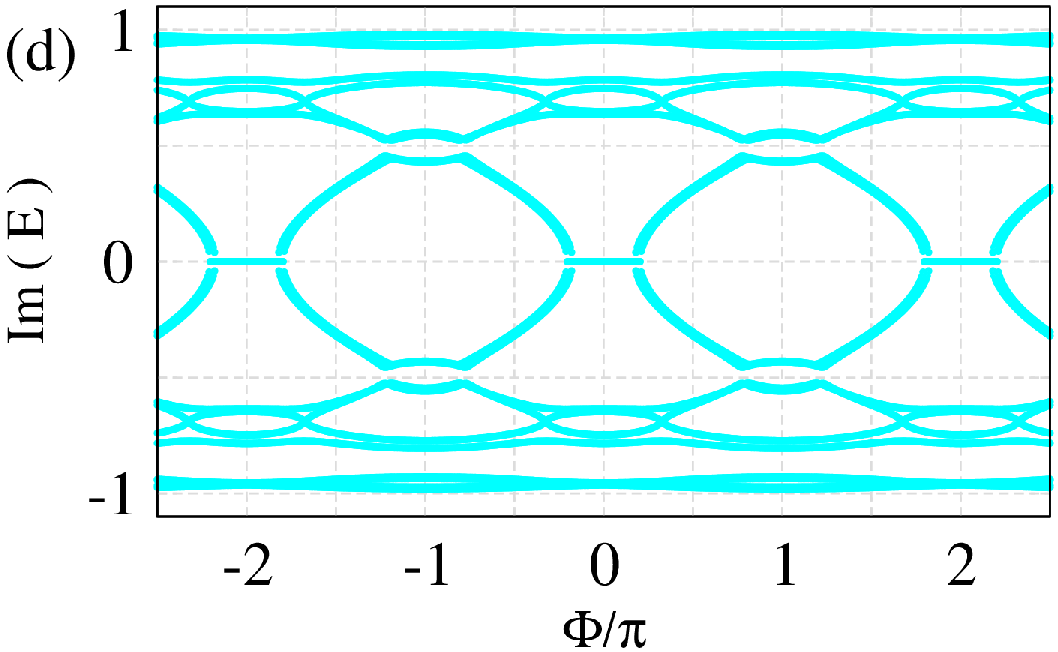}\hfill
\includegraphics[width=0.3\textwidth]{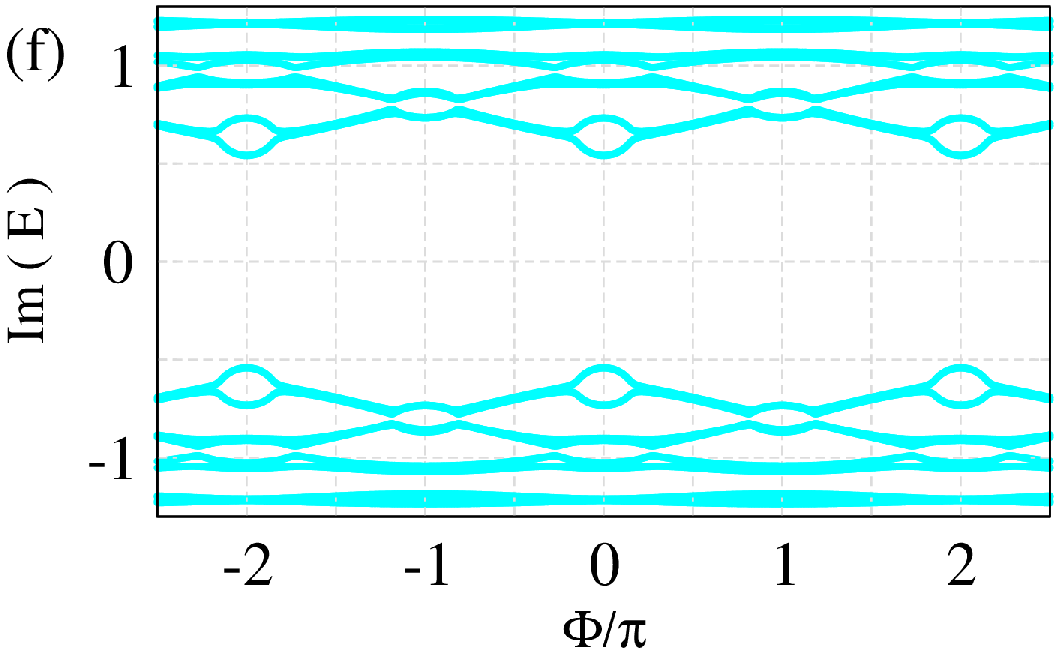}\vskip 0.1 in
\caption{(Color online.) Real and imaginary energy spectrum as a function of $\Phi$ in the presence of AAH disorder. (a) Real and (b) imaginary energies for
$t_1 = 0.75$. (c) Real and (d) imaginary energies for $t_1 = 1$. (e) Real and (f) imaginary energies for $t_1 = 1.25$. Number of unit cells $N = 10$ and intercell hopping strength $t_2 = 1$. The AAH disorder strength is fixed at $W=0.5$. An inset is shown in (a), where the lower half of the spectrum is zoomed in for clarity.}
\label{aah-energy}
\end{figure*}
The filling factor, defined as the ratio between the number of electrons present in the system and the total number of electronic states, plays a crucial role in determining the behavior of persistent currents. To analyze this, we compute the currents as a function of the filling factor, as illustrated in Fig.~\ref{fill}(a) for the real part and for the imaginary part in Fig.~\ref{fill}(b). The color conventions are the same as in Fig.~\ref{size}. The current profiles for all cases are symmetric about the half-filled case. This symmetry arises because the energy levels are symmetrically distributed around zero energy, as seen in Fig.~\ref{ephi}. For the given system size ($2N=80$), it is observed in the topological phase that the real current is zero (Fig.~\ref{size}(a)), while the imaginary current is zero in the trivial phase (Fig.\ref{size}(b)). These observations were made for the half-filled case, 
which is also evident from Figs.~\ref{fill}(a) and \ref{fill}(b). Interestingly, this is not the case for other filling factors. For example, in the window between the quarter-filled and half-filled cases in Fig.~\ref{fill}(a), the real current is not zero but has appreciable magnitudes (red curve). Similarly, for the imaginary part in the trivial phase (Fig.~\ref{fill}(b)), the green curve shows significant values. This clearly indicates that the current is highly sensitive to the filling factor. Moreover, the real (imaginary) current in the topological (trivial) phase attains its maximum value and gradually falls off on either side of the half-filled case. The same feature is also observed at the critical point $t_1=t_2$ as shown by the orange curves.

\subsection{Presence of disorder}
\subsubsection{AAH model}
With a clear understanding of the persistent current in a non-Hermitian ring and a few unexplored features, we now turn our attention to the interplay between non-Hermiticity and disorder effects. To begin, we consider the AAH model and examine the behavior of the energy spectrum, as depicted in Fig.~\ref{aah-energy}. For this study, we fix the number of unit cells to $N=10$ and the disorder strength at $W=0.5$. 

\begin{figure*}[ht]
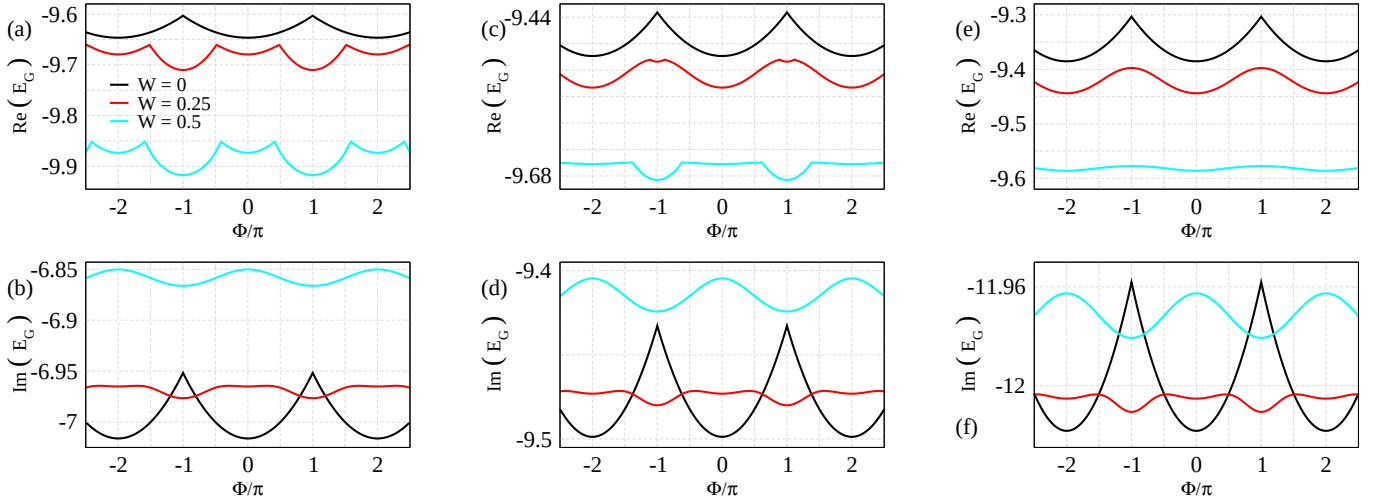
 
\includegraphics[width=0.3\textwidth]{fig8a.eps}\hfill
\includegraphics[width=0.3\textwidth]{fig8c.eps}\hfill
\includegraphics[width=0.3\textwidth]{fig8e.eps}\vskip 0.1 in
\includegraphics[width=0.3\textwidth]{fig8b.eps}\hfill
\includegraphics[width=0.3\textwidth]{fig8d.eps}\hfill
\includegraphics[width=0.3\textwidth]{fig8f.eps}
\caption{(Color online.) Ground state energy $E_G$ as a function of $\Phi$ in the presence of AAH disorder. (a) Real and (b) imaginary $E_G$ for $t_1 = 0.75$. (c) Real and (d) imaginary $E_G$ for $t_1 = 1$. (e) Real and (f) imaginary $E_G$ for $t_1 = 1.25$. Number of unit cells $N = 20$ and intercell hopping strength $t_2 = 1$. The AAH disorder strength $W=0,0.25$, and 0.5 and the corresponding results are denoted with black, red, and cyan colors, respectively. Number of electrons is fixed at $N_e = 10$, which is the quarter-filled case.}
\label{aah-eg}
\end{figure*}
\begin{figure*}[ht] 
\includegraphics[width=0.3\textwidth]{fig9a.eps}\hfill
\includegraphics[width=0.3\textwidth]{fig9c.eps}\hfill
\includegraphics[width=0.3\textwidth]{fig9e.eps}\vskip 0.1 in
\includegraphics[width=0.3\textwidth]{fig9b.eps}\hfill
\includegraphics[width=0.3\textwidth]{fig9d.eps}\hfill
\includegraphics[width=0.3\textwidth]{fig9f.eps}
\caption{(Color online.) Persistent current $I$ as a function of $\Phi$ in the presence of AAH disorder. (a) Real and (b) imaginary $I$ for $t_1 = 0.75$. (c) Real and (d) imaginary $I$ for $t_1 = 1$. (e) Real and (f) imaginary $I$ for $t_1 = 1.25$. Number of unit cells $N = 20$ and intercell hopping strength $t_2 = 1$. The AAH disorder strength $W=0,0.25$, and 0.5 and the corresponding results are denoted with black, red, and cyan colors, respectively. Number of electrons is fixed at $N_e = 10$, which is the quarter-filled case.}
\label{aah-i}
\end{figure*}

In the topological phase, the real energy spectrum is significantly altered in the presence of AAH disorder, as depicted in Fig.~\ref{aah-energy}(a). Notably, the gap around zero energy remains intact despite the disorder. 
The inset of Fig.~\ref{aah-energy}(a), showing the lower half of the real spectrum, reveals the presence of loop-like structures around $\Phi = 2n\pi$ ($n = 0, 1, 2, \ldots$). The imaginary energy spectrum also experiences significant changes, as shown in Fig.~\ref{aah-energy}(b), with the continuous evolution of imaginary eigenvalues completely disrupted by AAH disorder. An interesting observation is that these loop-like structures in the real spectrum persist within specific $\Phi$-windows, and within these same $\Phi$-windows, a few zero imaginary energies emerge. In both the real and imaginary spectra, the $\Phi$ periodicity remains $2\pi$.

At the critical point, the previously observed identical behavior of the real and imaginary components of the energy spectrum (disorder-free case) is disrupted due to the presence of AAH disorder, as depicted in Figs.~\ref{aah-energy}(c) and (d), respectively. Despite this, both spectra retain some of their characteristic features as a function of $\Phi$, similar to their behavior in the topological phase. The real energy spectrum shows a noticeable reduction in the gap, while the imaginary energy spectrum becomes completely gapless. Additionally, loop-like structures persist in the real energy spectrum around the specific $\Phi$ values identified earlier. Around these $\Phi$ values, the imaginary energy spectrum also exhibits similar behavior to that observed in the topological phase.

In the trivial phase, the real and imaginary energy spectra exhibit similar characteristics as in the absence of AAH disorder, as shown in Figs.~\ref{aah-energy}(e) and (f), respectively. The real energies evolve almost continuously, while the imaginary energy spectrum is characterized by a gap. Interestingly, the imaginary spectrum now displays loop-like structures at the same $\Phi$-values previously noted for the real spectra in the topological phase. Overall, the presence of disorder breaks the symmetry of the real spectra around zero energy, while the imaginary components remain consistently symmetric.

To compute the ground state energy $E_G$ in the presence of AAH disorder, we consider a relatively larger ring size with the number of unit cells set to $N = 20$ and fix the number of electrons at $N_e = 10$, corresponding to the quarter-filled case. The behavior of $E_G$ as a function of $\Phi$ is shown in Fig.~\ref{aah-eg} for different values of $t_1$ under various disorder strengths. Specifically, we consider $W = 0$, $0.25$, and $0.5$ for illustration purposes, with their corresponding results represented in black, red, and cyan colors, respectively. The real part of $E_G$ decreases at a particular disorder strength as we move from the topological phase (Fig.~\ref{aah-eg}(a)) to the critical point (Fig.~\ref{aah-eg}(c)) and further to the trivial phase (Fig.~\ref{aah-eg}(e)). In contrast, the imaginary part of $E_G$ increases as we progress from the topological phase (Fig.~\ref{aah-eg}(b)) to the critical point (Fig.~\ref{aah-eg}(d)) and finally to the trivial phase (Fig.~\ref{aah-eg}(f)). This is physically understandable. As the intradimer hopping integral $t_1$ increases, the non-Hermiticity of the system becomes more pronounced, resulting in larger imaginary eigenenergies. The real part of $E_G$ in the topological phase exhibits a distinct feature in the presence of disorder. Within the $\Phi$-window between $-1$ and $1$, the slope of the real ground state energy changes sign only once under clean conditions, but with disorder, the sign change occurs twice. At the critical point, this behavior is somewhat smoothed out in the presence of disorder, and in the trivial phase, the feature disappears entirely, becoming sinusoidal as a function of $\Phi$. Apart from differences in magnitude, the imaginary part of $E_G$ behaves in a similar manner across the three different regions when plotted as a function of $\Phi$ (Figs.~\ref{aah-eg}(b), (d), and (f)), given a fixed disorder strength. The similarity in behavior implies that, although the AAH disorder strength introduces complexity, the primary features influencing the imaginary part of $E_G$ are largely governed by the non-Hermiticity of the system across all regions.

\begin{figure*}[ht]
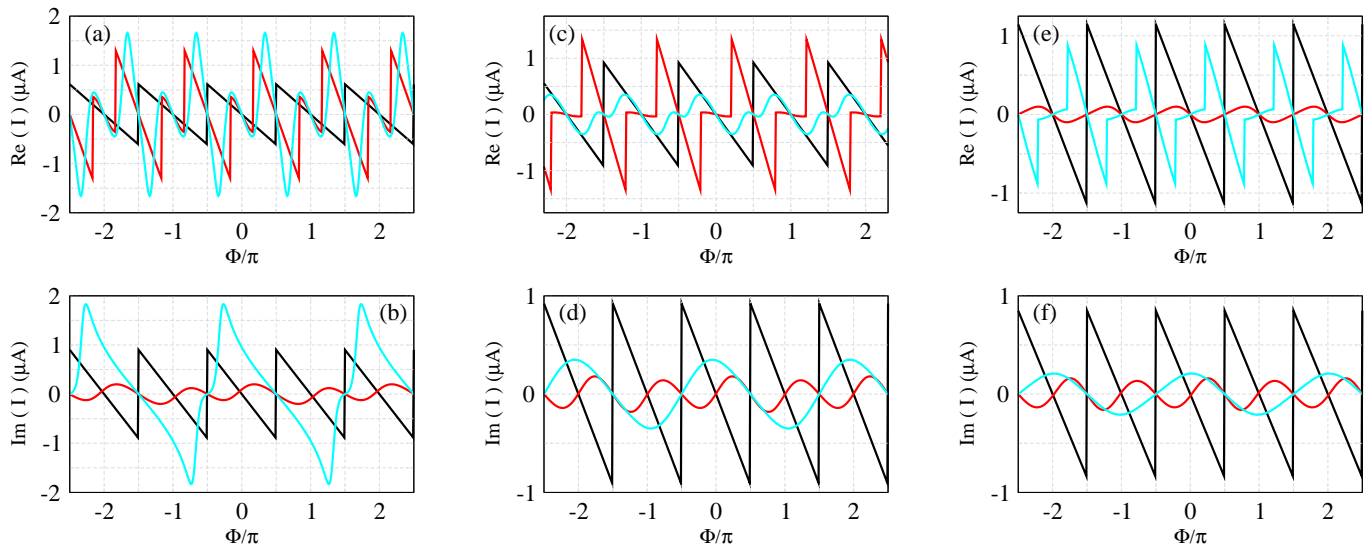
 
\includegraphics[width=0.3\textwidth]{fig10a.eps}\hfill
\includegraphics[width=0.3\textwidth]{fig10c.eps}\hfill
\includegraphics[width=0.3\textwidth]{fig10e.eps}\vskip 0.1 in
\includegraphics[width=0.3\textwidth]{fig10b.eps}\hfill
\includegraphics[width=0.3\textwidth]{fig10d.eps}\hfill
\includegraphics[width=0.3\textwidth]{fig10f.eps}
\caption{(Color online.) Persistent current $I$ as a function of $\Phi$ in the presence of Fibonacci disorder. (a) Real and (b) imaginary $I$ for $t_1 = 0.75$. (c) Real and (d) imaginary $I$ for $t_1 = 1$. (e) Real and (f) imaginary $I$ for $t_1 = 1.25$. The intercell hopping strength $t_2 = 1$. The AAH disorder strength $W=0,0.25$, and 0.5 and the corresponding results are denoted with black, red, and cyan colors, respectively. Number of unit cells is $N=17$ and the number of electrons is fixed at $N_e = 8$, which is close to the quarter-filled case.}
\label{fb-i}
\end{figure*}
The behavior of the persistent current as a function of $\Phi$ in the presence of AAH disorder is depicted in Fig.~\ref{aah-i}. All the system parameters, disorder strengths, electron filling factor, and color conventions are the same as those described in Fig.~\ref{aah-eg} and are also mentioned in the caption of Fig.~\ref{aah-i}. The real persistent current in the topological phase exhibits two distinct anomalous features, as shown in Fig.\ref{aah-i}(a). First, in the presence of AAH disorder with $W=0.25$, the real persistent current (red curve) is higher than in the disorder-free case (black curve), and with $W=0.5$, the current increases further compared to $W=0.25$. This indicates that increasing disorder strength enhances the current amplitude, which is unexpected. Second, while the current typically changes sign only once within the usual flux period of $2\pi$, the presence of AAH disorder causes the sign change to occur twice. This behavior is consistent with the ground state energy plot discussed in Fig.\ref{aah-eg}. Both of these features are highly atypical for disordered ring systems. At the critical point, the current behavior changes from the previous scenario. For $W=0.25$, the current is reduced compared to the disorder-free case, as shown in Fig.~\ref{aah-i}(c). However, as disorder strength increases to $W=0.5$, the current not only recovers but also exceeds the maximum value observed without disorder, demonstrating that current amplification remains significant even at the critical point. Although the sign change in the current is less pronounced than before, it is still present, indicating that disorder continues to influence the phase of the current, albeit more subtly. In contrast, the current amplification effect is entirely absent in the trivial phase, as depicted in Fig.~\ref{aah-i}(e). Here, the real current decreases systematically with increasing disorder strength. In the absence of disorder, the current exhibits a sawtooth-like behavior, but with disorder, it becomes sinusoidal as a function of $\Phi$, without any additional sign changes in the current. The most important observation from the above analysis is that AAH disorder leads to a favorable transport, as discussed in recent works~\cite{longi,cc-wan}.

The imaginary persistent current, in the absence of disorder, exhibits a sawtooth-like behavior across all three regions: the topological phase (Fig.~\ref{aah-i}(b)), the critical point (Fig.~\ref{aah-i}(d)), and the trivial phase (Fig.~\ref{aah-i}(f)). In terms of magnitude, it remains largely unchanged. In the presence of AAH disorder, while the magnitude of the imaginary current consistently decreases with increasing disorder strength, the overall behavior remains similar in all three regions, showing no significant qualitative changes.

\begin{figure*}[ht]
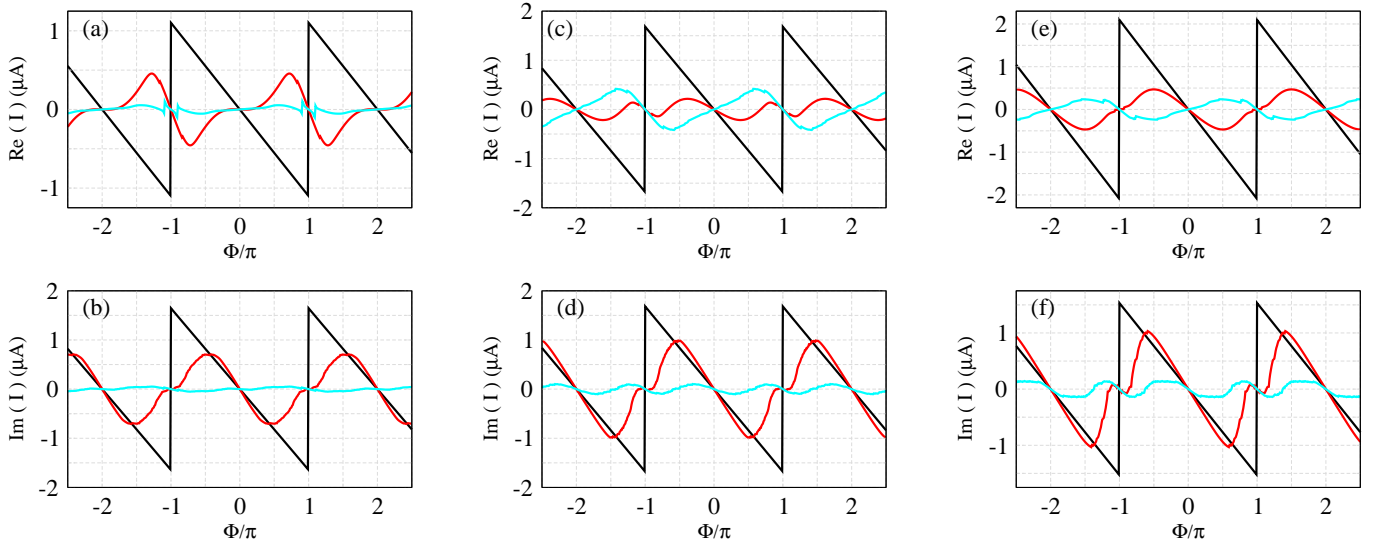
 
\includegraphics[width=0.3\textwidth]{fig11a.eps}\hfill
\includegraphics[width=0.3\textwidth]{fig11c.eps}\hfill
\includegraphics[width=0.3\textwidth]{fig11e.eps}\vskip 0.1 in
\includegraphics[width=0.3\textwidth]{fig11b.eps}\hfill
\includegraphics[width=0.3\textwidth]{fig11d.eps}\hfill
\includegraphics[width=0.3\textwidth]{fig11f.eps}
\caption{(Color online.) Persistent current $I$ as a function of $\Phi$ in the presence of random disorder. (a) Real and (b) imaginary $I$ for $t_1 = 0.75$. (c) Real and (d) imaginary $I$ for $t_1 = 1$. (e) Real and (f) imaginary $I$ for $t_1 = 1.25$. The intercell hopping strength $t_2 = 1$. The AAH disorder strength $W=0,0.25$, and 0.5 and the corresponding results are denoted with black, red, and cyan colors, respectively. Number of unit cells is $N=20$ and the number of electrons is fixed at $N_e = 10$, which is the quarter-filled case. All the currents are averaged over 100 configuration.}
\label{random-i}
\end{figure*}

\subsubsection{Fibonacci model}
Next, we consider correlated disorder following the Fibonacci sequence and examine the behavior of both the real and imaginary components of the persistent current as a function of magnetic flux, $\Phi$, as depicted in Fig.~\ref{fb-i}. In this case, the number of unit cells is fixed at $N = 17$, corresponding to 34 total sites—a value from the Fibonacci sequence. The number of electrons is set to $N_e = 8$, which is close to the quarter-filling case. The remaining parameters, including disorder strengths and color conventions, are the same as in Fig.~\ref{aah-i}. The behavior of the real persistent current in the topological phase follows a similar pattern to that observed under AAH disorder, as shown in Fig.~\ref{fb-i}(a). Specifically, the current increases as the strength of Fibonacci (FB) disorder increases. A key distinction, however, is that in the FB case, the current changes sign four times, whereas for AAH disorder, the sign reversal occurs only twice. At the critical point, the real current at $W = 0.25$ (red curve) is greater than in the absence of disorder (black curve). However, as the disorder strength continues to increase, the current diminishes, as depicted in Fig.~\ref{fb-i}(c). At $W = 0.5$ (cyan curve), the current reaches its lowest value, which is lower than both the disorder-free case and the $W = 0.25$ scenario. Interestingly, the number of sign changes at $W = 0.5$ remains the same as in the topological phase. In the trivial phase, the current at $W = 0.25$ is lower than in the disorder-free case, but it reaches a maximum at $W = 0.5$, as shown in Fig.~\ref{fb-i}(e). However, in contrast to the topological phase, the current only changes sign twice in the trivial phase. 

The amplification of the current is also evident in the behavior of the imaginary persistent current in the topological region, as shown in Fig.~\ref{fb-i}(b). In this case, the current at $W = 0.25$ is lower than in the disorder-free case, but interestingly, it reaches its maximum at $W = 0.5$. At the critical point, as seen in Fig.~\ref{fb-i}(d), the imaginary current is highest when no disorder is present, but the current at $W = 0.5$ still exceeds that at $W = 0.25$. In the trivial phase, depicted in Fig.~\ref{fb-i}(f), the imaginary current follows the typical trend observed with increasing disorder strength, steadily decreasing as the disorder strength grows.

\subsubsection{Random disorder}
This study also investigates the effect of random disorder on persistent currents. The variation of the current as a function of $\Phi$ in the presence of random disorder is illustrated in Fig.~\ref{random-i}. The system, consisting of $N = 20$ unit cells (as in the AAH model), is set to a quarter-filled case, that is $N_e = 10$. Currents are determined by averaging over 100 random disorder configurations. In contrast to correlated disorder, both the real and imaginary components of the persistent current exhibit a characteristic decreasing trend with increasing disorder strength across all scenarios.

It is essential to note that, although the averaged currents consistently show this decreasing trend, they reflect an ensemble average over all random configurations. A closer examination of individual random configurations reveals that a subset does exhibit current amplification. Nonetheless, as the majority of configurations align with the conventional decreasing trend, and given that persistent currents from different configurations also vary in sign at a fixed $\Phi$, the overall averaged effect remains a net decrease in the persistent current under random disorder. The behavior of persistent currents for a representative configuration is additionally illustrated in Fig.~\ref{s1} of the supporting material~\cite{support}.

\begin{figure}[h] 
\includegraphics[width=0.48\textwidth]{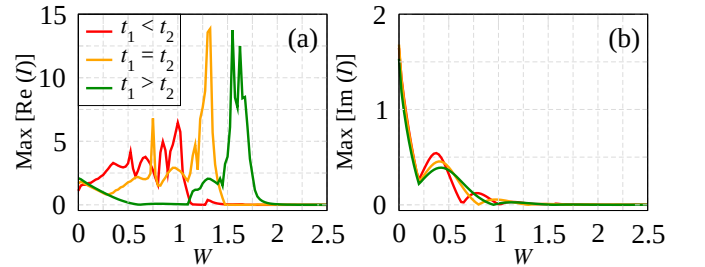}
\caption{(Color online.) $\text{Max}[I]$ as a function of AAH disorder strength. (a) Real and (b) imaginary currents. The number of unit cells is $N=20$ and number of electrons is fixed at $N_e=10$. Definition of $\text{Max}[I]$ is described in the texts. Red, orange, and green colors represent the result corresponding to the topological phase, critical point, and trivial phase, respectively.}
\label{aah-var}
\end{figure}

To gain a comprehensive understanding of the impact of disorder on the persistent current, we vary the disorder strength over a complete flux period, ranging from $-\pi$ to $\pi$, and focus on the maximum amplitude of the current, denoted as $\text{Max}[I]$. The behavior of $\text{Max}[I]$ as a function of AAH disorder strength is shown in Figs.~\ref{aah-var}(a) and (b) for the real and imaginary components, respectively.

In Fig.~\ref{aah-var}(a), the real persistent current exhibits intriguing behavior in response to disorder. In the topological phase, the current increases with disorder strength, showing fluctuations, reaches a maximum at $W = 1$, and then drops sharply, becoming nearly zero beyond this point. At the critical point, the current initially decreases with disorder strength up to $W \sim 0.25$, after which it begins to increase with fluctuations and reaches 
its highest value around $W \approx 1.25$. However, beyond this threshold, the current drops abruptly to zero. A similar pattern is observed in the trivial phase, where the maximum current occurs at a higher disorder strength compared to the other two cases. This trend indicates that the critical disorder strength, beyond which the real current becomes vanishingly small, increases with the intradimer hopping integral $t_1$.

This behavior is physically intuitive. As $t_1$ increases, the kinetic energy of the electrons also increases, requiring a higher critical disorder strength to fully localize the system. Thus, a stronger disorder is needed to suppress the current when the electrons are more mobile.

On the other hand, the imaginary persistent current follows a different pattern, as shown in Fig.~\ref{aah-var}(b). It decreases steadily with increasing disorder strength up to $W \approx 0.25$, where it shows a small hump across all the phases. Beyond this point, the current decreases again as $W$ continues to increase. Overall, current amplification is more pronounced in the real part compared to the imaginary part in the presence of AAH disorder.

\begin{figure}[h] 
\includegraphics[width=0.48\textwidth]{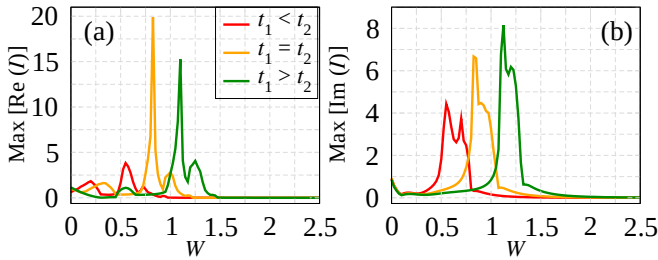}
\caption{(Color online.) $\text{Max}[I]$ as a function of Fibonacci disorder strength. (a) Real and (b) imaginary currents. The number of unit cells is $N=17$ and number of electrons is fixed at $N_e=8$. Definition of $\text{Max}[I]$ is described as in Fig.~\ref{aah-var}. Red, orange, and green colors represent the result corresponding to the topological phase, critical point, and trivial phase, respectively.}
\label{fb-var}
\end{figure}

The behavior of persistent current with FB disorder strength is presented in Fig.~\ref{fb-var}, where the number of unit cells is set at $N = 17$ and the number of electrons is fixed at $N_e = 8$. The definition of $\text{Max}[I]$ is the same as in Fig.~\ref{aah-var}, and the color 
conventions remain consistent. For the real persistent current, a similar trend to that observed in the AAH case is seen, as shown in Fig.~\ref{fb-var}(a). The critical disorder strength, beyond which the real current becomes vanishingly small, increases with the intradimer hopping integral, just as in the AAH case. This indicates that higher disorder strength is required to fully localize the system as electron mobility increases with higher intradimer hopping. The imaginary persistent current exhibits a distinct and interesting pattern across the three phases. Initially, in all phases, the imaginary current decreases with increasing disorder strength, $W$. However, as $W$ continues to rise, the current begins to increase, reaches a maximum, and then sharply drops to zero due to the onset of localization. Overall, in the presence of FB disorder, both the real and imaginary currents exhibit current amplification.

\begin{figure*}[ht]
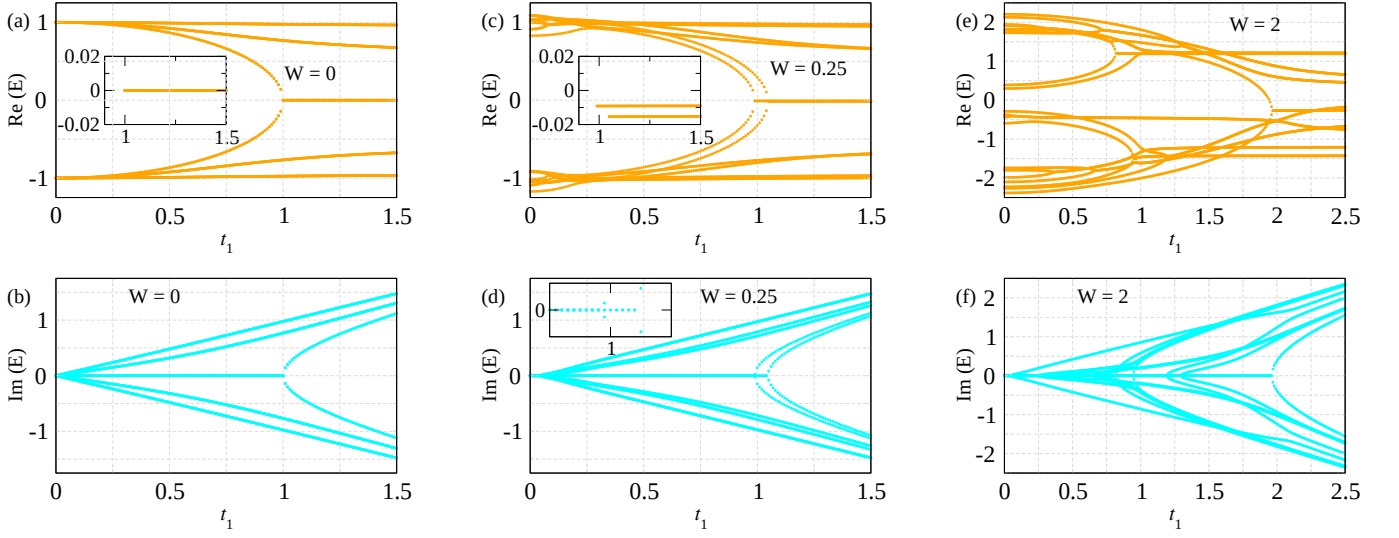
 
\includegraphics[width=0.3\textwidth]{fig14a.eps}\hfill
\includegraphics[width=0.3\textwidth]{fig14c.eps}\hfill
\includegraphics[width=0.3\textwidth]{fig14e.eps}\vskip 0.1 in
\includegraphics[width=0.3\textwidth]{fig14b.eps}\hfill
\includegraphics[width=0.3\textwidth]{fig14d.eps}\hfill
\includegraphics[width=0.3\textwidth]{fig14f.eps}
\caption{(Color online.) Energy spectrum as a function of intradimer hopping integral $t_1$ in the presence of AAH disorder. (a) Real and (b) imaginary for $W=0$. (c) Real and (d) imaginary $E$ for $W=0.25$. (e) Real and (f) imaginary $E$ for $W=2$. The interdimer hopping integral is fixed at $t_2=1$ and magnetic flux at $\Phi=0$. Number of unit cells $N = 10$.}
\label{t1-var}
\end{figure*}

{\it A note on critical point}: So far, we have defined the critical point as the condition $t_1=t_2$. However, in the presence of disorder, as shown in Fig.~\ref{aah-energy}, the identical behavior between the real and imaginary parts of the energy spectrum is disrupted. Nonetheless, both the real and imaginary components retain their line-gap properties in the topological and trivial phases, respectively. Interestingly, at the critical point, the real line-gap structure is preserved (Fig.~\ref{aah-energy}(c)). This raises an important question regarding the validity of the critical point under disorder: should the condition $t_1=t_2$ still be considered a critical point in the presence of disorder? Alternatively, does the critical point shift to different values of the hopping integrals, or is it altogether destroyed, implying no such critical point remains under disorder? To resolve this issue, we study the behavior of real and imaginary energies as a function of the intradimer hopping integral $t_1$ as shown in Fig.~\ref{t1-var}. 

We begin by discussing the behavior of real and imaginary energies as a function of $t_1$ for the disorder-free case, as illustrated in Figs.~\ref{t1-var}(a) and (b), respectively. In Fig.~\ref{t1-var}(a), the real energy spectrum reveals a gap in the range $t_1=0$ to 1, beyond which a zero-energy line emerges. This is further validated by the inset in Fig.~\ref{t1-var}(a), where it is evident that precisely at $t_1=1$, a zero-energy line forms. For the imaginary part, shown in Fig.~\ref{t1-var}(b), a zero-energy line persists up to $t_1=1$. Beyond this point, a gap appears.

Introducing disorder at $W=0.25$, we observe that the real energy spectrum continues to display a gap up to $t_1=1$, followed by the emergence of a zero-energy line. However, as shown in the inset of Fig.~\ref{t1-var}(c), no zero-energy line exists beyond $t_1=1$ within the given range of intradimer hopping integrals. In the case of imaginary energies (Fig.~\ref{t1-var}(d)), and as highlighted in the inset, the gap opens not exactly at $t_1=1$ but at a slightly higher value.

For a larger disorder strength, $W=2$, the real energy spectrum clearly lacks a zero-energy line, as shown in Fig.~\ref{t1-var}(e). The imaginary energy spectrum in Fig.~\ref{t1-var}(f) further reveals that the gap opens around $t_1=2$. Thus, the presence of disorder disrupts the real energy gap entirely, while the imaginary gap persists but shifts to higher values of $t_1$ as disorder increases. The greater the disorder, the higher the intradimer hopping integral required for the imaginary line gap to appear. It can be emphasized that the features of critical point disappear in the presence of disorder.

{\it Bond-resolved analysis of persistent currents}:
It is also interesting to study the distribution of persistent current across the bonds. For that, we adopt the current operator method. The current operator is defined as~\cite{maiti-physicae}
\begin{equation}
\hat{J} = \frac{e\dot{X}}{2Na},
\end{equation}
where $e$ is the electronic charge, $\dot{X}$ is the velocity operator, $2N$ is the number of sites, and $a$ is the lattice spacing. The velocity operator is further defined as
\begin{equation}
\dot{X} = \frac{2\pi}{ih}[X,H],
\end{equation}
where $X$ is the position operator and $H$ is the system Hamiltonian as in Eq.~\ref{ham}. The position operator can be expressed in terms of the fermionic operators as
\begin{equation}
X = \sum_{n=1}^N na\left(c_{n,A}^{\dagger} c_{n,A} + c_{n,B}^{\dagger} c_{n,B}\right).
\end{equation}

By taking the commutator between $X$ and $H$, the current operator gets the following form
\begin{eqnarray}
\hat{J} &=& \frac{2\pi e \lvert t_1 \rvert}{i (2N) h} \sum_{n=1}^N \left(e^{i\phi} c_{n,A}^{\dagger} c_{n,B} + e^{-i\phi} c_{n,B}^{\dagger} c_{n,A}\right) \nonumber\\
&+& \frac{2\pi e t_2}{i (2N) h} \sum_{n=1}^{N-1} \left(c_{n,B}^{\dagger} c_{n+1,A} - c_{n+1,A}^{\dagger} c_{n,B}\right) \nonumber\\
&+& \frac{2\pi e t_2}{i (2N) h} \left(c_{N,B}^{\dagger} c_{1,A} - c_{1,A}^{\dagger} c_{N,B}\right).
\end{eqnarray}

For an $m$-th eigenstate $\lvert\psi_m\rangle$, the persistent current can be calculated as
\begin{equation}
I_m = \langle \psi_m\lvert \hat{J}\rvert\psi_m\rangle,
\end{equation}
where $\lvert\psi_m\rangle = \sum_p \left(u_{p,A}^m\lvert p, A\rangle + u_{p,B}^m\lvert p, B\rangle\right)$. Here $\lvert p, A\rangle$ and $\lvert p, B\rangle$ are the Wannier states and $u_{p,A}^m$ and $u_{p,B}^m$ are the corresponding coefficients. The current for the $m$-th eigenstate takes the form
\begin{eqnarray}
I_m &=& \frac{2\pi e \lvert t_1 \rvert}{i (2N) h}\sum_{p=1}^N \left[e^{i\phi}\left(u_{p,A}^m\right)^* u_{p,B}^m + e^{-i\phi}\left(u_{p,B}^m\right)^* u_{p,A}^m   \right]\nonumber\\
&+&\frac{2\pi e t_2}{i (2N) h}\sum_{p=1}^{N-1} \left[\left(u_{p,B}^m\right)^* u_{p+1,A}^m - \left(u_{p+1,A}^m\right)^* u_{p,B}^m  \right]\nonumber\\
&+&\frac{2\pi e t_2}{i (2N) h} \left[\left(u_{N,B}^m\right)^* u_{1,A}^m - \left(u_{1,A}^m\right)^* u_{N,B}^m  \right].
\label{wf-cur}
\end{eqnarray}

The net persistent current for the occupied electronic states will be 
\begin{equation}
I = \sum_m I_m.
\label{tot-cur}
\end{equation}

\begin{figure}[h]
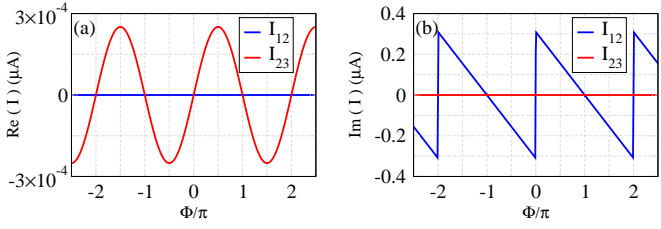
 
\includegraphics[width=0.23\textwidth]{fig15a.eps}\hfill\includegraphics[width=0.23\textwidth]{fig15b.eps}
\caption{(Color online.) Bond currents $I_{12}$ and $I_{23}$ as a function of $\phi$. (a) Real and (b) imaginary currents in the topological phase with $t_1=0.75$ and $t_2=1$. The number of unit cells is $N=20$ and number of electrons is fixed at $N_e=20$. The red and blue colors denote the real and imaginary currents, respectively.}
\label{bond-cur}
\end{figure}

The real and imaginary parts of Eq.~\ref{tot-cur} can be extracted to compute the real and imaginary components of the persistent current. Notably, both the derivative method (outlined in Sec.~\ref{sec2}) and the operator method produce identical results. However, the latter method offers the advantage of calculating the current across different bonds directly from Eq.\ref{wf-cur}. The current through any intradimer bond (from site $p$ to $q$) for the occupied electronic states can be written as
\begin{equation}
I_{pq} = 
\frac{2\pi e \lvert t_1 \rvert}{i (2N) h}\sum_{m} \left[e^{i\phi}\left(u_{p,A}^m\right)^* u_{q,B}^m + e^{-i\phi}\left(u_{q,B}^m\right)^* u_{p,A}^m\right],
\label{intra-cur}
\end{equation} 
while, that through any inter-dimer bond for the occupied electronic states is
\begin{equation}
I_{pq} = \frac{2\pi e t_2}{i (2N) h}\sum_{m} \left[\left(u_{p,B}^m\right)^* u_{q,A}^m - \left(u_{q,A}^m\right)^* u_{p,B}^m\right].
\label{inter-cur}
\end{equation} 
A careful inspection of Eqs.~\ref{intra-cur} and \ref{inter-cur} reveals a unique property of the bond-resolved persistent current. In Eq.\ref{inter-cur}, the second term within the parentheses is the complex conjugate of the first term. Since the amplitudes $u_{p,A}^m$ and $e^{i\phi}$ are generally complex, the expression inside the parentheses becomes a real quantity. Consequently, with the factor $i$ outside the parentheses, the current through the intra-dimer bond is always imaginary. Similarly, for the inter-dimer bond, the current is always a real quantity. This behavior persists both in the absence and presence of disorder and is illustrated in Fig.~\ref{bond-cur} within the topological phase. The bond currents are plotted as a function of $\Phi$ for the half-filled case, focusing on the intra-dimer bond between sites 1 and 2 ($I_{12}$) and the inter-dimer bond between sites 2 and 3 ($I_{23}$). The real parts are shown in Fig.~\ref{bond-cur}(a), while the imaginary parts are depicted in Fig.\ref{bond-cur}(b). 
All system parameters match those used in Fig.~\ref{ip}(a). The real and imaginary currents are represented by blue and red lines, respectively. Consistent with the bond current expressions, the real part of the current is identically zero for the intradimer bond in Fig.\ref{bond-cur}(a), while a finite imaginary current appears in Fig.~\ref{bond-cur}(b). Conversely, the interdimer bond exhibits the opposite behavior. Furthermore, we observe that all intradimer bonds carry identical imaginary currents, while all inter-dimer bonds carry identical real currents. However, for the sake of brevity, this result is not explicitly shown in the present work. Thus, in such NH systems, real and imaginary bond currents appear alternately which is in complete contrast to the Hermitian ones.

The topological invariant of the Hermitian SSH model is characterized by the winding number, which takes a value of unity in the topological phase and zero in the trivial phase~\cite{short-course}. Due to the {\it bulk-boundary correspondence}, the topological phase hosts two localized zero-energy edge states, whereas the trivial phase does not support any such states. While the winding number is computed under periodic boundary condition, the existence of edge states must be examined under open boundary condition.

\begin{figure}[ht]
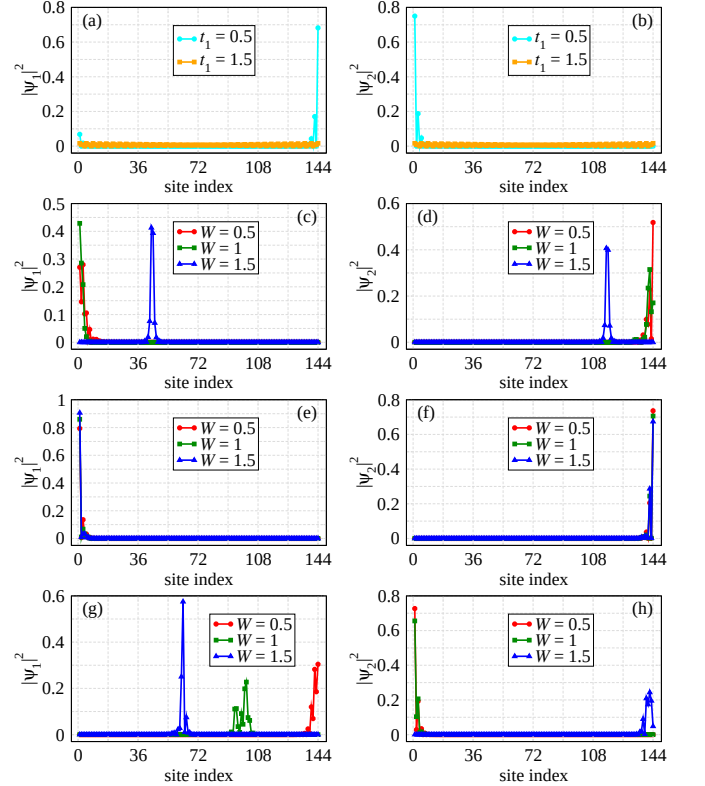
 
\includegraphics[width=0.235\textwidth]{fig16a.eps}\hfill
\includegraphics[width=0.235\textwidth]{fig16b.eps}\\
\includegraphics[width=0.235\textwidth]{fig16c.eps}\hfill
\includegraphics[width=0.235\textwidth]{fig16d.eps}\\
\includegraphics[width=0.235\textwidth]{fig16e.eps}\hfill
\includegraphics[width=0.235\textwidth]{fig16f.eps}
\includegraphics[width=0.235\textwidth]{fig16g.eps}\hfill
\includegraphics[width=0.235\textwidth]{fig16h.eps}
\caption{(Color online.) Distribution of local probability amplitude as a function of site index for edge modes $\psi_1$ and $\psi_2$ in a 1D NH chain with 144 sites. (a) and (b) In the absence of disorder, (c) and (d) in the presence of AAH disorder, (e) and (f) in the presence of FB disorder, and (g) and (h) in the presence of random disorder. The interdimer hopping is fixed as $t_2=1\,$eV in all the plots. In (c-h), only the topological phase is considered where the intradimer hopping is $t_1=0.5\,$eV.}
\label{edge-state}
\end{figure}

{\it Open boundary condition}:
To gain deeper insight, it is important to analyze the localization behavior of these edge states under open boundary condition. Specifically, we investigate their properties both in the clean limit (absence of disorder) and in the presence of three different types of disorder. To do so, we plot the probability amplitude as a function of site index as shown in Fig.~\ref{edge-state}.

\begin{table*}
\caption{Key features of the real and imaginary energy spectra, edge states, and real and imaginary persistent currents of HN ring without and with different disorder types in the topological and trivial phases.}
\label{table}
\begin{center}
\begin{tabular}{ |p{0.16\textwidth} | p{0.27\textwidth} | p{0.27\textwidth} | p{0.27\textwidth} | }
\hline
 \RaggedRight{Disorder type} & Energy spectrum & Edge states & Persistent current \\\hline

 \RaggedRight{Absence of disorder} & $\bullet$ Real part: Gapped in the topological phase.\newline
$\bullet$ Imaginary part: Gapped in the trivial phase.\newline
$\bullet$ Zero modes appear in the real gap. & $\bullet$ Two zero modes exist in the topological phase, localized at opposite edges.\newline
$\bullet$ In the trivial phase, these modes become extended. & $\bullet$ Both real and imaginary components exist.\newline
$\bullet$ The real component dominates in the trivial phase.\newline
$\bullet$ The imaginary component dominates in the topological phase.\newline
$\bullet$ The smaller component does not vanish completely. \\\hline

 \RaggedRight{AAH disorder} & $\bullet$ Real part: Gapped in the topological phase, but becomes fragmented with disorder.\newline
$\bullet$ Imaginary part: Gapped in the trivial phase, with disorder-induced broadening. & $\bullet$ Edge states persist at moderate disorder ($W=0.5, 1$).\newline
$\bullet$ For strong disorder ($W=1.5$), they become bulk-localized. & $\bullet$ Enhances both real and imaginary persistent currents compared to the disorder-free case.\newline
$\bullet$ Disorder can amplify persistent current before strong localization sets in. \\\hline

 \RaggedRight{Fibonacci} &  $\bullet$ Real part: Gapped in the topological phase, with disorder-induced fragmentation, same as AAH\newline
$\bullet$ Imaginary part: Gapped in the trivial phase, with disorder-induced broadening, same as AAH. & $\bullet$ Edge states remain localized at the edges for all disorder strengths.\newline
$\bullet$ Unlike AAH disorder, there is no sharp critical point for bulk localization. & $\bullet$ Similar enhancement of real and imaginary currents as AAH disorder.\newline
$\bullet$ Persistent current remains robust under strong disorder. \\\hline

 \RaggedRight{Uncorrelated (random) disorder} &  $\bullet$ Real part: Gap is robust at weak disorder strengths in the topological phase, highly configuration-dependent.\newline
$\bullet$ Imaginary part: Gap is robust at weak disorder strengths in the trivial phase, fluctuates across disorder realizations. & $\bullet$ Edge states remain localized at weak disorder ($W=0.5$).\newline
$\bullet$ At moderate to strong disorder ($W=1, 1.5$), they become bulk-localized with no systematic trend. & $\bullet$ Some individual disorder configurations show an increase in persistent current.\newline
$\bullet$ However, after averaging, this amplification disappears.\newline
$\bullet$ The imaginary component exhibits irregular variations across disorder realizations. \\\hline
\end{tabular}
\end{center}
\end{table*}

As mentioned earlier, under open boundary condition, a Hermitian SSH chain always hosts two zero modes in the topological phase, which are localized at the edges. Similarly, in the non-Hermitian case, such edge states persist, as illustrated in Figs.~\ref{edge-state}(a) and (b), where $\psi_1$ and $\psi_2$ represent the two edge states of a chain with 144 sites. The interdimer hopping integral is set to $t_2 = 1\,$eV. Results corresponding to the topological phase are shown in cyan, while those for the trivial phase are shown in orange. In Fig.~\ref{edge-state}(a), one of the zero modes, denoted as $\psi_1$, is localized at the right edge of the non-Hermitian SSH chain, whereas the other zero mode, $\psi_2$, is localized at the left edge, as shown in Fig.~\ref{edge-state}(b). In the trivial phase, the zero modes are extended throughout the chain. The topological and trivial phases for the non-Hermitian chain can be characterized by the winding number. Calculations show that in the topological phase the winding number is 1 and in the trivial phase, it is zero~\cite{nrh8,dip}. 

The effect of disorder on localized edge states is examined for three different disorder 
strengths, namely $W = 0.5$, 1, and 1.5, across the three types of disorder. In the AAH case, the modes $\psi_1$ and $\psi_2$ remain localized at either edge of the chain for $W = 0.5$ and 1. However, at a higher disorder strength of $W = 1.5$, they eventually become localized in the bulk, as shown in Figs.~\ref{edge-state}(c) and (d). Thus, the localized edge modes persist up to a moderate AAH strength before becoming bulk-localized with increasing disorder. In an AAH-disordered chain, these localized edge states are topologically protected, where the topology of the systems can be mapped to the lattice version of the 2D integer quantum Hall effect (IQHE)~\cite{kraus}. An interesting feature emerges in the Fibonacci case that is, both states, $\psi_1$ and $\psi_2$, remain localized regardless of the disorder strength, as seen in Figs.~\ref{edge-state}(e) and (f). These localized edge states are also topologically protected and are equivalent to the edge states of the 2D IQHE~\cite{verb}. The key distinction between the AAH and Fibonacci models lies in their criticality, while the former becomes critical at a sharply defined point in parameter space, the latter remains critical irrespective of the modulation strength~\cite{anu,anna,zij}. For the uncorrelated disordered chain, we consider a specific random configuration of the on-site energies. For $W=0.5$, the modes $\psi_1$ and $\psi_2$ remain localized at the edges, as seen from Figs.~\ref{edge-state}(g) and (h). As the disorder strength increases, the behavior of these modes changes. $\psi_1$ becomes localized in the bulk for $W=1$ and 1.5, while $\psi_2$ remains an edge-localized mode at $W=1$ before also becoming bulk-localized at $W=1.5$. A thorough inspection of different disorder configurations confirms that the edge modes always remain localized at the edges for $W=0.5$, but beyond this, no systematic localization trend is observed.

It should be noted that in the presence of disorder, translational invariance is lost, making it challenging to compute topological invariants from the momentum-space Hamiltonian. However, in recent years, an alternative approach has emerged, focusing on the computation of real-space winding numbers in both Hermitian~\cite{winding-h1, winding-h2, winding-h3} and non-Hermitian systems~\cite{winding-nh4}. Several recent studies have specifically explored the real-space chiral winding number in disordered non-Hermitian systems~\cite{winding-nh1, winding-nh2, winding-nh3}. The computation of the real-space winding number is feasible only if the system preserves chiral symmetry. In Hermitian disordered systems, these methods yield an exactly quantized winding number, enabling the characterization of distinct topological phases even in the presence of strong disorder. Similarly, in non-Hermitian systems, the winding number remains quantized and has been shown to be closely related to the non-Hermitian skin effect. However, since our system does not preserve the chiral symmetry in the presence of disorder, it is not possible for us to compute the real-space winding number in the present work. Nevertheless, we find that localized edge states persist in the regime of weak to moderate disorder strength, similar to topological edge states in the non-Hermitian SSH model. Their survival is driven by the interplay between dimerized hopping integrals and disorder.


For clarity, in Table~\ref{table}, we provide a summary of the key findings and notable differences in the real and imaginary energy spectra, the behavior of real and imaginary persistent currents, and the edge states under open boundary condition, both in the absence and presence of different disorder types. The energy spectra for the Fibonacci case and uncorrelated case are depicted in the supporting information~\cite{support} of Figs.~\ref{s2} and \ref{s3}, respectively. 

{\bf Experimental feasibility:} To realize an electronic Hatano-Nelson (HN) ring with anti-Hermitian intradimer hopping and disordered on-site potentials, one requires a mesoscopic quantum ring where electron transport can be engineered with controlled non-reciprocity and disorder. Quantum rings have already been fabricated, and persistent currents in such systems have been measured using superconducting quantum interference devices (SQUID)~\cite{pcex1, pcex2} by detecting the magnetization of the ring geometry. In our proposed setup, while the real component of the persistent current can be inferred from the magnetization of the system, the imaginary component may be estimated by measuring the rate of change of magnetization.

However, incorporating anti-Hermitian intradimer hopping alongside disorder presents a significant experimental challenge.

The fabrication of the quantum ring structure can be achieved using molecular beam epitaxy (MBE)~\cite{qd-ex1, qd-ex2}, which allows for precise control over the size and shape of individual quantum dots (QDs) forming the ring. Disorder can be introduced in the system by modifying the on-site potentials of individual QDs. This can be implemented by applying selective voltages to local electrostatic gates near each QD. By carefully tuning these gate voltages, one can generate specific potential landscapes such as the AAH or Fibonacci disorder models. If the gate voltages are randomized, an uncorrelated disorder potential can be realized.

For the realization of anti-Hermitian intradimer hopping, the key requirement is to engineer asymmetric tunneling amplitudes between certain QD pairs. Specifically, one needs hopping terms of the form $t + i\gamma$ from QD-1 to QD-2 and $-t + i\gamma$ from QD-2 to QD-1. The real staggered hopping ($\pm t$) can be introduced by employing tunable electrostatic gates between alternate QD pairs to create staggered barrier potentials, enforcing the required sign alternation in the hopping matrix elements.

The imaginary component of the hopping, $i\gamma$, can be introduced by coupling alternate QDs to electron reservoirs. By adjusting the coupling strength between a QD and its attached reservoir, one can regulate electron leakage from the QD, thereby controlling the effective magnitude of $\gamma$. 

Thus, by integrating controlled disorder through gate-defined on-site potentials and non-reciprocal hopping via QD-reservoir coupling, an experimental realization of an anti-Hermitian HN ring with disorder becomes feasible in mesoscopic electronic systems.

\section{\label{conclusion}Summary}
This work investigates the interplay of topology, diagonal disorder, and non-Hermiticity in the behavior of persistent current within the Hatano-Nelson ring. Non-Hermiticity is introduced through the anti-Hermitian intradimer hopping integral, which acts as a synthetic magnetic field. We consider disorder using two widely studied correlated disorder models, namely the AAH model and the FB model. For completeness, we also examine uncorrelated (random) disorder. Our analysis encompasses the energy spectrum, ground state energy, and the persistent current, including both real and imaginary components, under varying condition in the absence and presence of disorder. 

In the absence of disorder, we uncover that the behavior of persistent current strongly depends on the topology of the system and filling factor. In the topological (trivial) phase, the real (imaginary) persistent current is strictly zero, as indicated in recent work~\cite{nrh8}. However, this statement is only valid for larger ring systems in the half-filled case. The revised conclusion we draw is that in the topological (trivial) phase, the imaginary (real) persistent current is the dominant one, while the real (imaginary) current remains weak. As the system size increases, both the real and imaginary currents decrease toward zero. Even in larger ring systems, the topological (trivial) phase may host a real (imaginary) persistent current for specific filling factors, other than the half-filled case.

The effect of AAH disorder introduces complexity into the real and imaginary energy spectra while preserving the primary features observed in the absence of disorder. In the topological phase, a real line gap is present, while the trivial phase hosts an imaginary line gap. Notably, the real and imaginary energy spectra lose their identical behavior characteristic of the disorder-free scenario.

In the topological phase, the real persistent current exhibits an intriguing amplification phenomenon as the disorder strength increases, particularly at the critical point. In contrast, the imaginary persistent current follows a conventional decreasing trend with increasing disorder.

On the other hand, FB disorder leads to current amplification in both the real and imaginary persistent currents. In the case of uncorrelated disorder, no atypical behavior is observed overall. Interestingly, some individual uncorrelated configurations do exhibit current amplification, which, however, disappears when averaged over multiple configurations, indicating that while individual realizations may show enhanced current, the average effect does not sustain this phenomenon.

The observation from the analysis of bond-resolved currents is that the complex persistent current in the ring flows as alternating real and imaginary currents through the bonds. Specifically, the intra-dimer bonds exclusively host imaginary currents, while the inter-dimer bonds carry only real currents.

By analyzing the localization behavior of local probability amplitudes, we find that edge states remain localized under weak to moderate disorder strengths, similar to topological edge states in the non-Hermitian SSH model, for both the AAH model and random disorder. In contrast, for the Fibonacci model, localized edge states remain robust regardless of disorder strength. Their persistence stems from the interplay between dimerized hopping amplitudes and disorder effects.

Overall, our findings illustrate the intricate interplay among the disorder, hopping dimerization, and non-Hermiticity in various topological phases, emphasizing the importance of disorder type and configuration in determining the behavior of the persistent currents. Additionally, we have outlined an experimental proposal to realize and investigate these effects in a physical setup. Therefore, the exploration of these effects not only enriches our understanding of non-Hermitian quantum systems but also suggests potential avenues for future research into disorder-induced phenomena.

\setcounter{secnumdepth}{0}

\pagebreak
\widetext
\begin{center}
\textbf{\large Supplemental Materials\vskip 0.1 in Persistent current in a non-Hermitian Hatano-Nelson ring: Disorder-induced amplification}
\end{center}
\setcounter{equation}{0}
\setcounter{figure}{0}
\setcounter{table}{0}
\setcounter{page}{1}
\makeatletter
\renewcommand{\theequation}{S\arabic{equation}}
\renewcommand{\thefigure}{S\arabic{figure}}
\renewcommand{\bibnumfmt}[1]{[S#1]}
\renewcommand{\citenumfont}[1]{S#1}

\begin{figure}[h]
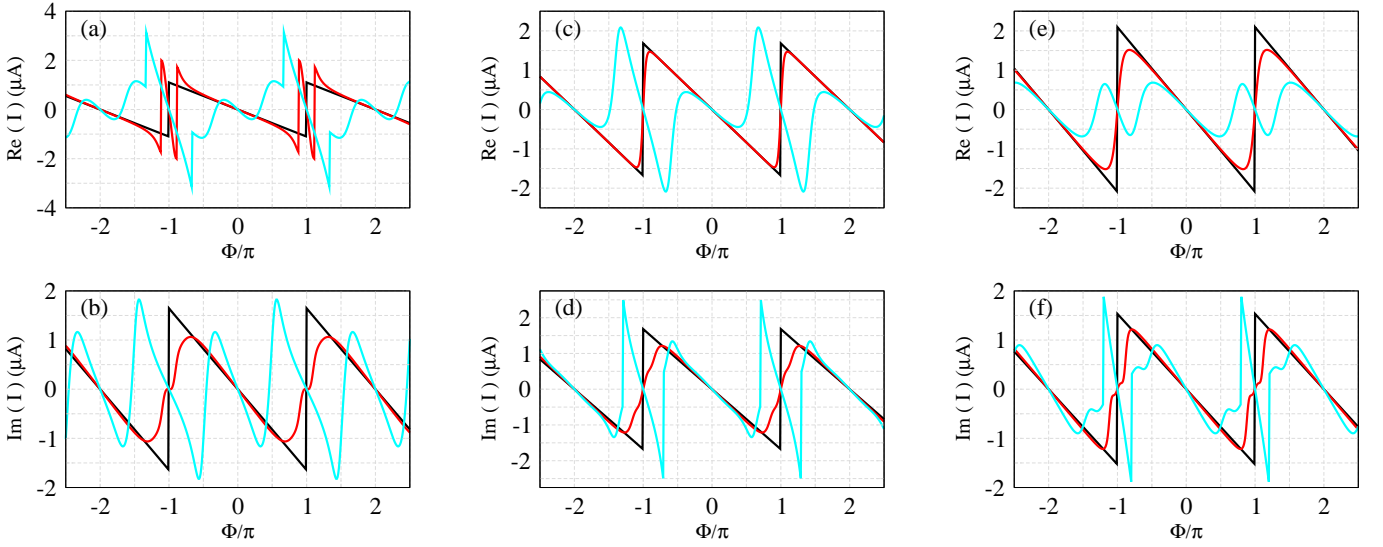

\centering
\includegraphics[width=0.3\textwidth]{sid237_t1_0.75_real.eps}\hfill\includegraphics[width=0.3\textwidth]{sid237_t1_1_real.eps}\hfill\includegraphics[width=0.3\textwidth]{sid237_t1_1.25_real.eps}\vskip 0.1 in
\includegraphics[width=0.3\textwidth]{sid237_t1_0.75_imag.eps}\hfill\includegraphics[width=0.3\textwidth]{sid237_t1_1_imag.eps}\hfill\includegraphics[width=0.3\textwidth]{sid237_t1_1.25_imag.eps}
\caption{(Color online.) Persistent current $I$ as a function of $\Phi$ in the presence of random disorder for a single configuration. (a) Real and (b) imaginary $I$ for $t_1 = 0.75$. (c) Real and (d) imaginary $I$ for $t_1 = 1$. (e) Real and (f) imaginary $I$ for $t_1 = 1.25$. The interdimer hopping strength $t_2 = 1$. Disorder strengths $W=0,0.25$, and 0.5 and the corresponding results are denoted with black, red, and cyan colors, respectively. Number of unit cells is $N=20$ and the number of electrons is fixed at $N_e = 10$, which is the quarter-filled case.}
\label{s1}
\end{figure}

Figure~\ref{s1} illustrates the behavior of the real and imaginary components of the persistent current as a function of $\Phi$ for a specific random configuration. Here, the current amplification is evident in both components across different phase values, showcasing deviations from the average trend observed in the ensemble analysis in the main text.

\begin{figure}[h]
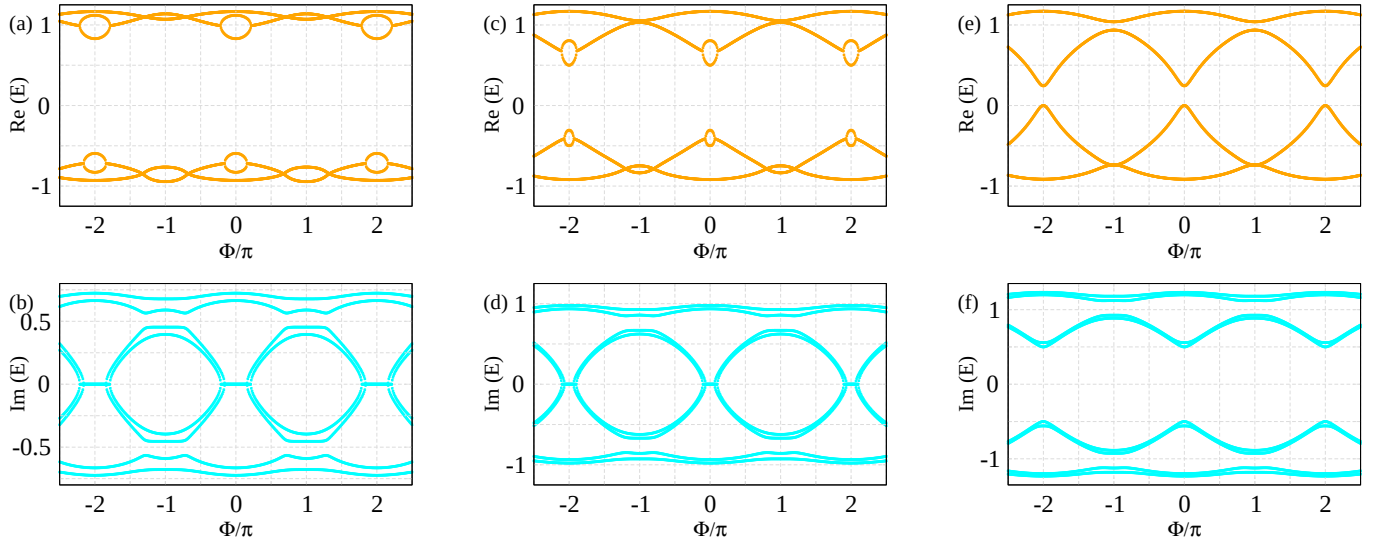

\centering
\includegraphics[width=0.3\textwidth]{real_eng_topo.eps}\hfill\includegraphics[width=0.3\textwidth]{real_eng_critical.eps}\hfill\includegraphics[width=0.3\textwidth]{real_eng_trivial.eps}\vskip 0.1 in
\includegraphics[width=0.3\textwidth]{img_eng_topo.eps}\hfill\includegraphics[width=0.3\textwidth]{img_eng_critical.eps}\hfill\includegraphics[width=0.3\textwidth]{img_eng_trivial.eps}
\caption{(Color online.) Real and imaginary energy eigenspectra as a function of $\Phi$, where the first, second and, third columns are for $t_1=0.75$, $1$, and $1.25$, respectively. The Fibonacci disorder strength is $W=0.5$. Here, the number of sites 8 and the interdimer hopping strength $t_2=1$.}
\label{s2}
\end{figure}

In connection with Table~I in the main text, we present the energy spectra for the Fibonacci-disordered case in Fig.~\ref{s2}. The system size is fixed at 8 sites, with an interdimer hopping of $t_2=1\,$eV and a disorder strength of $W=0.5$. Notably, the characteristic energy spectrum observed in the absence of disorder remains largely preserved at this disorder strength.
The real energy spectrum in the topological phase exhibits a clear gap, as shown in Fig.~\ref{s2}(a). A similar trend is observed at the critical point (Fig.~\ref{s2}(c)), albeit with a reduced gap compared to the topological phase. In the trivial phase, the band gap is further diminished, as illustrated in Fig.~\ref{s2}(e). Although fragmentation is not prominent in this small system, it becomes more visible for larger rings across all phases. For the imaginary part of the spectrum, a gap is present in the trivial phase (Fig.~\ref{s2}(f)), whereas no such gap appears in the topological phase (Fig.~\ref{s2}(b)) or at the critical point (Fig.~\ref{s2}(d)).

\begin{figure}[h]
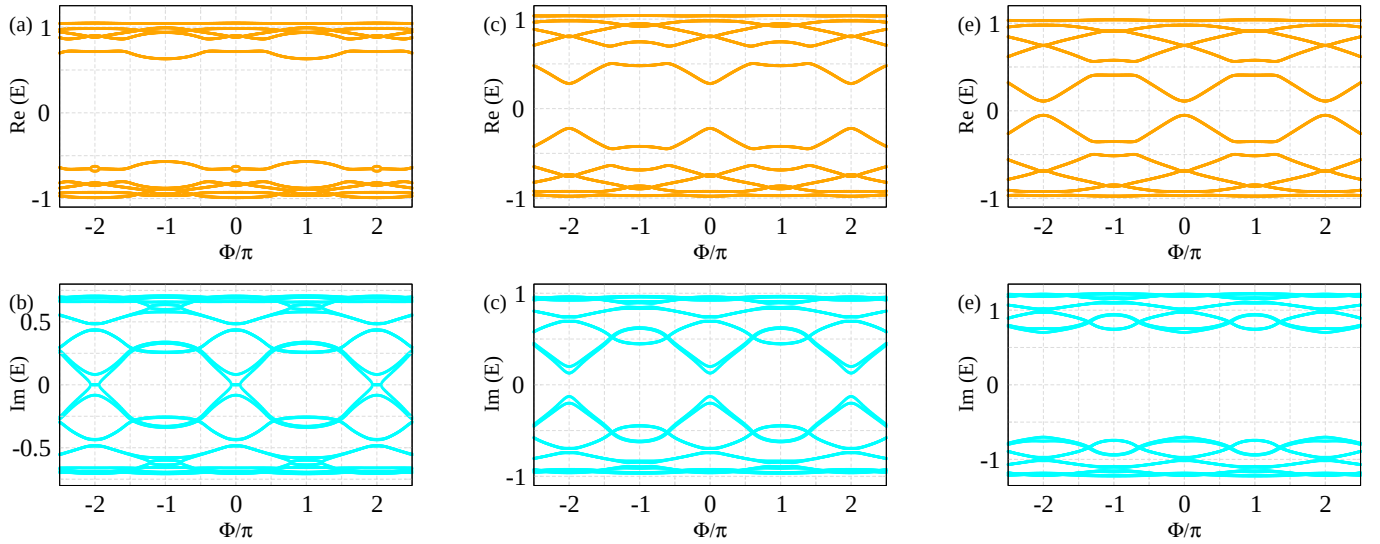

\centering
\includegraphics[width=0.3\textwidth]{rand_real_eng_topo.eps}\hfill\includegraphics[width=0.3\textwidth]{rand_real_eng_critical.eps}\hfill\includegraphics[width=0.3\textwidth]{rand_real_eng_trivial.eps}\vskip 0.1 in
\includegraphics[width=0.3\textwidth]{rand_img_eng_topo.eps}\hfill\includegraphics[width=0.3\textwidth]{rand_img_eng_critical.eps}\hfill\includegraphics[width=0.3\textwidth]{rand_img_eng_trivial.eps}
\caption{(Color online.) Real and imaginary energy eigenspectra as a function of $\Phi$, where the first, second and, third columns are for $t_1=0.75$, $1$, and $1.25$, respectively, for a specific random configuration. The disorder strength is $W=0.5$. Here, the number of unit cells $N= 20$ and the interdimer hopping strength $t_2=1$.}
\label{s3}
\end{figure}

The behavior of the energy spectrum with $\Phi$ for the uncorrelated disorder case is shown in Fig.~\ref{s3} for a specific configuration. Here, the random disorder strength is fixed at $W = 0.5$, and the number of unit cells is set to $N = 10$. The interdimer hopping remains the same as in the previous cases, i.e., $t_2 = 1\,$eV. Interestingly, the real energy spectrum exhibits a gap in the topological phase. We have verified this across numerous other configurations and consistently found the same result, suggesting that the gap is robust at weak disorder. However, in the trivial phase, the real spectrum may or may not exhibit a gap, depending on the specific configuration. For the imaginary spectrum in the trivial phase, a gap is present at weak disorder strength, and this feature appears to be robust, similar to the real part.

\end{document}